\documentclass[aps,prb,floatfix,superscriptaddress,notitlepage]{revtex4-1}
\usepackage{bm}
\usepackage{amsmath,amssymb,graphicx}
\usepackage{bbold}
\usepackage{braket,titlesec,natbib}
\usepackage{dsfont}
\usepackage{comment}
\usepackage{float}
\usepackage{color}
\usepackage{hyperref}
\usepackage{graphicx}
\usepackage[capitalize]{cleveref} 
\hypersetup{colorlinks=true,linkcolor=blue,anchorcolor=blue,citecolor=blue,filecolor=blue,urlcolor=blue,bookmarksnumbered=true,pdfview=FitB}

\newcommand{\bse}{\begin{subequations}}
\newcommand{\ese}{\end{subequations}}
\newcommand{\re}{\text{Re}}

\newcommand{\bxi}{\hat{\boldsymbol{\xi}}}
\newcommand{\btau}{\hat{\boldsymbol{\tau}}}

\newcommand{\bv}{{\bf v}}

\newcommand{\beq}{\begin{equation}}
\newcommand{\eeq}{\end{equation}}
\newcommand{\bea}{\begin{eqnarray}}
\newcommand{\eea}{\end{eqnarray}}
\newcommand{\ve}{\varepsilon}

\newcommand{\bk}{{\bf k}}

\newcommand{\br}{{\bf r}}
\newcommand{\bu}{{\bf u}}
\newcommand{\bw}{{\bf w}}
\newcommand{\bE}{{\bf E}}

\newcommand{\nn}{\nonumber}
\newcommand{\bwt}{\begin{widetext}}
\newcommand{\ewt}{\end{widetext}}
\newcommand{\im}{\mathrm{Im}}

\newcommand{\hta}{\hat{\tau}}

\newcommand{\er}{\eqref}
\newcommand{\lr}{\lambda_\mathrm{R}}
\newcommand{\lz}{\lambda_\mathrm{Z}}
\newcommand{\vs}{\varsigma}

\begin{document}

\title{Zero-field spin resonance  in graphene\\ with proximity-induced spin-orbit coupling}

\author{Abhishek Kumar}
\affiliation{Center for Materials Theory, Department of Physics and Astronomy, Rutgers University, Piscataway, NJ 08854, USA}
\affiliation{Department of Physics, University of Florida, Gainesville, Florida, 32611, USA}
\author{Saurabh Maiti}
\affiliation{Department of Physics and Centre for Research in Molecular Modeling, Concordia University, Montreal, QC  H4B 1R6, Canada}
\author{Dmitrii L. Maslov}
\affiliation{Department of Physics, University of Florida, Gainesville, Florida, 32611, USA}
\date{\today}
\begin{abstract} We investigate collective spin excitations in graphene with proximity-induced spin-orbit coupling (SOC) of the Rashba and valley-Zeeman types, as it is the case, e.g.,  for graphene on transition-metal-dichalcogenide substrates. It is shown that, even in the absence of an external magnetic field,  such a system supports collective modes, which correspond to coupled oscillations of the uniform and valley-staggered magnetizations. These modes can be detected via both zero-field electron spin resonance (ESR) and zero-field electric-dipole spin resonance (EDSR), with EDSR response coming solely from Rashba  SOC. We analyze the effect of electron-electron interaction within the Fermi-liquid kinetic equation and show that the  interaction splits both the ESR and EDSR peaks into two. The magnitude of splitting and the relative weights of the resonances can be used to extract the spin-orbit coupling constants and many-body interaction parameters that may not be accessible by other methods.
\end{abstract}

\maketitle
\newpage
\section{Introduction}

A magnetic field applied to a system of interacting fermions gives rise to a collective mode: a Silin spin wave, \cite{silin1958,nozieres,statphys,baym} in which spins precess coherently around the direction of the magnetic field. (We will be referring to the magnetic field acting on electron spins as to ``Zeeman field''.)
It has been shown in a number of theoretical studies that a combination of SOC (of Rashba and Dresselhaus types) and electron-electron interaction leads to a new type of spin collective modes: chiral spin waves, which occur even in the absence of the external magnetic field.\cite{shekhter2005, ashrafi2012, ashrafi2013, zhang2013, maiti2014, 
maiti2016, 
kumar2017} 
If both the Zeeman field and SOC are present, the collective modes are of the mixed Silin/chiral wave type. Such mixed modes were observed by Raman spectroscopy in 2D semiconductor heterostructures in the regime when the magnetic field is stronger than SOC, \cite{perez:2013,perez:2015,perez:2016,karimi:2017} In the absence of the Zeeman field, a collective spin mode was  observed by Raman spectroscopy on the surface state of a three-dimensional (3D)  topological insulator.\cite{kung2017} The $q=0$ end points of the spin waves' spectra can be probed by electron spin resonance (ESR), if the mode is driven by an {\em ac} magnetic field, and by  electric-dipole spin resonance (EDSR), if the mode is driven by an {\em ac} electric field which couples to electron spins via spin-orbit interaction.\cite{rashba:1991,rashbaefros2003,efros2006} 
 
A free-standing graphene or graphene on a substrate made of light elements (SiO$_2$, hBN) can have only an intrinsic Kane-Mele (KM) type of SOC,\cite{KM} which is very weak.
However, a much stronger SOC can be induced in a graphene layer deposited on a heavy-metal substrate  and/or intercacalated with heavy-metal atoms, see, e.g., Ref.~\onlinecite{Ren:2016} and references therein. First-principle calculations and experiments have demonstrated that the proximity-induced Rashba SOC in graphene on heavy-metal substrates (Au, Ni, Pb, Ir, Co) can reach up to 100 meV.\cite{varykhalov:2008,Marchenko2012,Krivenkov:2017} Another very popular platform is monolayer and bilayer graphene on transition-metal-dichalcogenide (TMD) substrates, such as WS$_2$, WSe$_2$ and MoS$_2$.\cite{Avsar:2014,morpurgo_2015,morpurgo_2016,Yang:2016,shi:2017,Dankert:2017,Ghiasi:2017,wakamura:2018,schonenberger:2018,Omar:2018,Benitez:2018,wakamura:2019,Wang:2019,Island:2019} 
In this case, the induced SOC is expected to be a mixture of two types:\cite{morpurgo_2015,cummings:2017,garcia:2018} of Rashba SOC, which leads to in-plane spin-momentum textures, and of valley-Zeeman (VZ) or Ising  SOC, which acts as an out-of-plane magnetic field whose direction alternates between the $K$ and $K'$ valleys of graphene. EDSR in graphene with both Rashba and VZ types of SOC has been predicted in a recent theoretical study,\cite{Raines:2021b} in which the Zeeman field was assumed to be much stronger than both types of SOC.  In this case, the frequency of the modes is set by (renormalized) Zeeman energies in channels with different angular momenta, while spin-orbit interaction provides a means to couple the driving electric field to electron spins.

In this work, we study ESR and EDSR in doped graphene with Rashba and VZ types of SOC in the absence of the Zeeman field. In this case, the resonance frequencies are determined by spin-orbit energy scales, renormalized by the electron-electron interaction. By applying an extension of the Fermi-liquid (FL) theory to the case of two electron valleys,\cite{Raines:2021} we show that the eigenmodes of the system correspond to coupled oscillations of the uniform and valley-staggered magnetizations. The coupling between the two sectors is provided by VZ SOC. If the latter is absent, two sectors are decoupled, and both ESR and EDSR signals detect only the uniform magnetization mode. If both Rashba and VZ types of SOC are present, {each of} the ESR and EDSR signals is split into two peaks due to the coupling between the two sectors. This coupling is mediated by the combined effect of the electron-electron interaction in the spin- and spin-valley channels and of VZ SOC. Finally, if only VZ SOC is present, ESR detects only one mode while EDSR shows only a continuum due to transitions between the spin-split valence and conduction bands, with a threshold around $\omega\sim 2\mu$, where $\mu$ is the chemical potential. In the current literature, the relative strengths of Rashba and VZ components of SOC in graphene on TMD is still an open issue: while some studies indicate that Rashba SOC is the dominant one,\cite{morpurgo_2015,morpurgo_2016,Yang:2016,shi:2017,Omar:2018} others find a stronger VZ component.\cite{wakamura:2018,schonenberger:2018,wakamura:2019} We analyze how the ESR and EDSR spectra depend on the interplay between the two types of SOC and propose to use ESR and EDSR experiments as a direct way to resolve this controversy.

The rest of the paper is organized as follows. In Secs.~\ref{Sec:Hamiltonian}, \ref{sec:free_RSOC}, and \ref{sec:free_VZ}, we describe our model and discuss the selection rules for ESR and EDSR in the absence of electron-electron interaction.  In Sec.~\ref{Sec:spinX}, we derive a low-energy Hamiltonian for the conduction band. Section \ref{Sec:IntModel} deals with the effects of electron-electron interaction. In Sec.~\ref{sec:2VFL}, we introduce the two-valley FL theory. In Sec.~\ref{sec:eigen}, we discuss the eigenmodes of a two-valley FL with Rashba and VZ types of SOC. ESR and EDSR in this system are described in Secs.~\ref{sec:esr} and \ref{sec:edsr}, respectively. In Sec.\ref{sec:Conclusions}, we present our conclusions and discuss the feasibility of an experimental observation of the effects predicted in this paper. Technical details of the calculations are delegated to Appendices \ref{app:free} and \ref{Kin}.

\section{Electron spin and electric-dipole spin resonances in the non-interacting system}
\label{sec:free}
\subsection{Single-particle Hamiltonian}
\label{Sec:Hamiltonian}
We consider a monolayer graphene attached to a substrate made of, e.g,  TMD, or heavy metal. Strong SOC in the substrate induces SOC in graphene, which can be generically of both Rashba and VZ types. For completeness, we also allow for an intrinsic KM term.
Following Refs.~\onlinecite{KM,rashba:2009,stauber:2009,morpurgo_2015,cummings:2017,garcia:2018},
 we adopt the following low-energy Hamiltonian:
\bea
\label{ham0}
\hat{H}_0 &=& v_F(\tau_z \hat{s}_0 \hat{\sigma}_x k_x + \hat{s}_0 \hat{\sigma}_y k_y) + \Delta \hat{s}_0 \hat{\sigma}_z + \frac{\lambda_{\rm KM}}{2} \tau_z \hat{s}_z \hat{\sigma}_z +  \frac{\lambda_{\rm R}}{2} (\tau_z \hat{s}_y \hat{\sigma}_x - \hat{s}_x \hat{\sigma}_y) + \frac{\lambda_{\rm Z}}{2} \tau_z \hat{s}_z \hat{\sigma}_0,
\eea
where $v_F$ is the Dirac velocity; $\textbf{k}$ is the electron momentum measured either from the $K$ or $K'$ point of the graphene Brillouin zone, $\lambda_{\rm KM,Z,R}$ are the coupling constants of the KM, VZ, and Rashba spin-orbit interactions, respectively, $\Delta$ is the gap due to substrate-induced asymmetry between the A and B sites of the honeycomb lattice, $\hat{\sigma}_i$ and $\hat{s}_i$ are the Pauli matrices in the sublattice (pseudospin) and spin spaces, respectively (with $\hat \sigma_0$ and $\hat s_0$ being the unity matrices in the corresponding spaces), and $\tau_z=\pm 1$ labels the $K$ and $K'$ points. 

The first three terms in Eq.~\er{ham0} describe (massive) Dirac fermions near the $K$ and $K'$ points, the rest of the terms describe SOC.
The intrinsic SOC present locally on the sublattice sites of graphene gives rise to two terms in the Hamiltonian, 
which are symmetric and antisymmetric combinations of local SOCs on the A and B sites. Because free-standing graphene is invariant under sublattice exchange, only the symmetric combination, which was identified by Kane and Mele,\cite{KM} survives. It couples spins, sublattices (pseudo-spins), and valleys, as indicated by the fourth term in Eq.~\er{ham0}.  The presence of a substrate brings about new features. First, the breaking of $z\rightarrow-z$ inversion symmetry induces a Rashba-type SOC. At the $K$and $K'$ points, the Rashba Hamiltonian contains the leading, momentum-independent term [the fifth term in Eq.~\er{ham0}] and the subleading, linear-in-$k$ term.\cite{kane,rashba:2009}
We assume that doping is low enough, i.e., $k_F\ll\lambda_R/\alpha$, where $\alpha$ is the Rashba parameter for the linear-in-$k$ coupling, such that the latter can be neglected, and Rashba SOC will be described by the fifth term in Eq.~\er{ham0}. Second, if the substrate also breaks the sublattice symmetry,  the anti-symmetric combination of atomic SOCs gives rise to a VZ SOC--the last term in Eq.~\er{ham0}. 

The Hamiltonian in Eq.~\er{ham0} can be simplified further. First, it is well known that the KM coupling is much weaker than other types of SOC. While theoretical estimates place $\lambda_{\rm KM}$ in the range from 1\,$\mu$eV \cite{min:2006,yao:2007} to 25-50\,$\mu$eV,\cite{trickey:2007,konschuh:2010}
a recent ESR experiment reports the value of $42.2$\, $\mu$eV. \cite{sichau:2019} This is much smaller than typical values of 1-10 meV for $\lambda_{\rm R}$ and $\lambda_{\rm Z}$ for graphene on TMD substrates,  \cite{morpurgo_2015,morpurgo_2016,cummings:2017,wakamura:2018,schonenberger:2018,wakamura:2019}  and thus the KM term will be ignored in what follows. Next, substrate-induced sublattice asymmetry is expected to open a gap of magnitude $\Delta$ at the Dirac points. This gap endows graphene with Berry curvature which, in combination with SOC, leads to interesting consequences for EDSR, such as a Hall component of the induced current, as discussed recently in Ref.~\onlinecite{Raines:2021b}. In the regime of $\lr,\lz\ll \mu$ (where the FL theory of Secs.~\ref{sec:eigen}-\ref{sec:edsr} is valid), the presence of the gap gives rise to the Hall conductivity but does not qualitatively affect the collective modes, as it amounts only to changes in the Fermi velocity and SOC parameters. Given also that angular-resolved photoemission of graphene on TMD substrates has not detected gaps in the Dirac spectrum, \cite{Coy-Diaz:2015,Pierucci:2016,Henck:2018}  we will neglect the asymmetry gap in this study. With these simplifications, the Hamiltonian is reduced to
\bea
\label{ham1}
\hat{H}_0 &=& v_F(\tau_z \hat{s}_0 \hat{\sigma}_x k_x + \hat{s}_0 \hat{\sigma}_y k_y) + \frac{\lambda_{\rm Z}}{2} \tau_z \hat{s}_z \hat{\sigma}_0 + \frac{\lambda_{\rm R}}{2} (\tau_z \hat{s}_y \hat{\sigma}_x - \hat{s}_x \hat{\sigma}_y).
\eea

\subsection{Energy spectrum and selection rules for Rashba spin-orbit coupling}
\label{sec:free_RSOC}
If only Rashba SOC is present ($\lambda_{\rm Z}=0$), the eigenvalues and eigenstates of the Hamiltonian \er{ham1} are given by\cite{rashba:2009,stauber:2009} 
\beq
\begin{split}
\label{es}
\ve_{\alpha \beta;\tau_z}(\bk) &= \alpha\left( \sqrt{v_F^2 k^2+\left(\frac{\lambda_{\rm R}}{2}\right)^2} + \beta \frac{\lambda_{\rm R}}{2} \right);\\
\ket{\alpha \beta;\tau_z} &= \frac{1}{\sqrt{2(1+(\epsilon_{\alpha\beta})^{2\tau_z})}} 
\left( \begin{array}{c}
- i\alpha \beta \tau_z  e^{-i (1+\tau_z) \theta} \\
-i \alpha \beta e^{-i \theta} (\epsilon_{\alpha\beta})^{\tau_z} \\
\tau_z e^{-i \tau_z \theta} (\epsilon_{\alpha\beta})^{\tau_z}  \\ 
1
\end{array} \right),~\epsilon_{\alpha\beta} = \ve_{\alpha \beta;\tau_z}(\bk)/v_F k,
\end{split}
\eeq
where $\alpha = \pm 1$ denotes the conduction/valence band, $\beta = \pm 1$ denotes the SOC-split chiral subbands, $\tau_z=\pm1$ denotes the $K/K'$ valleys, respectively, and $\theta$ is the azimuthal angle of \textbf{k}.
\begin{figure}
\centering
\includegraphics[scale=0.5]{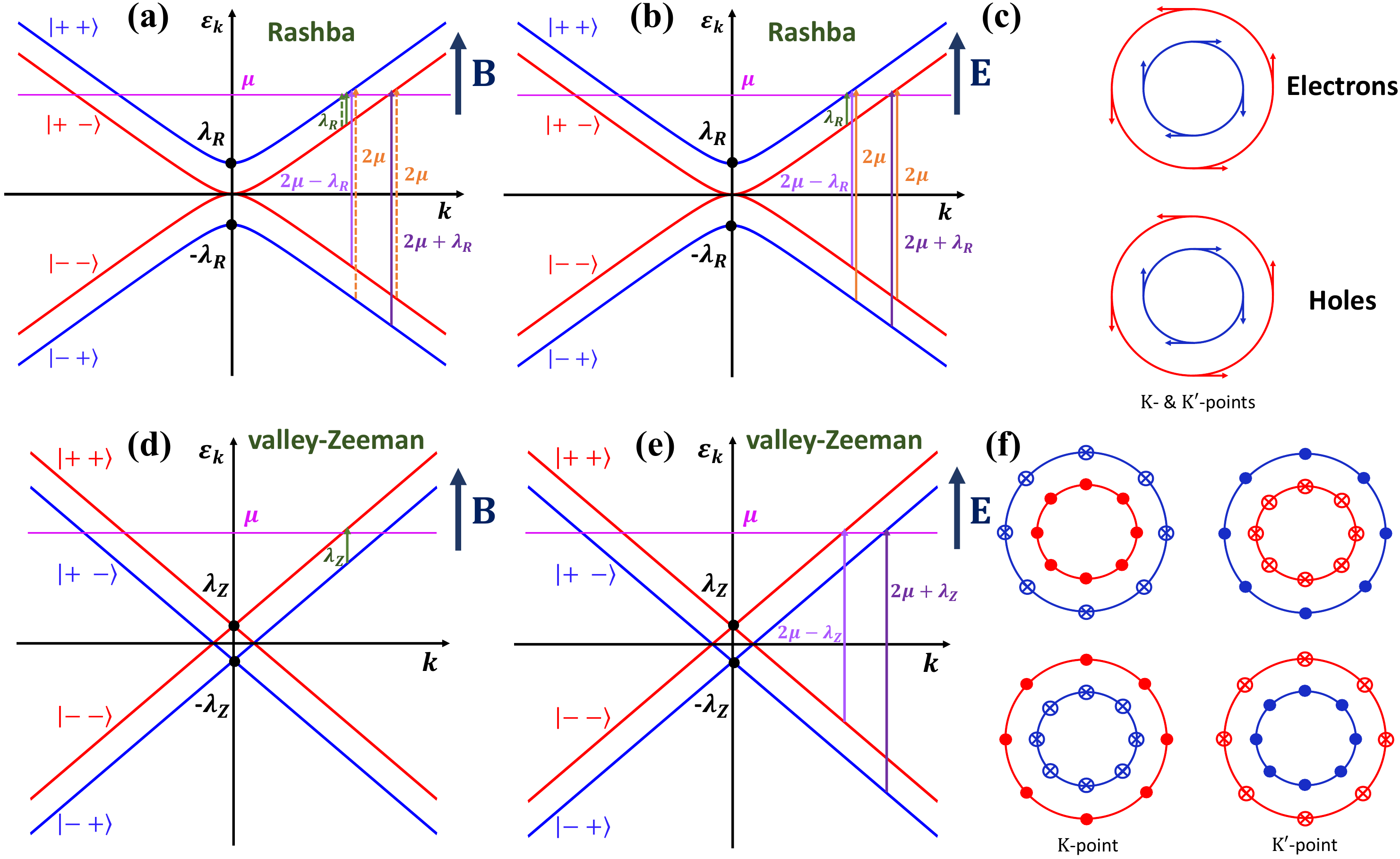}
\caption{\label{dispersionR}(a) Energy spectrum of graphene with Rashba spin-orbit coupling for $\lambda_\mathrm{R}<\mu$. The blue (red) lines depict subband dispersions with negative (positive) chirality, in the right-handed notation. Vertical arrows indicate transitions induced by an {\em ac} magnetic field. Solid and dashed arrows show transitions induced by an in-plane and out-of-plane magnetic field, respectively. b) Same spectrum as in (a) with arrows showing transitions induced by an {\em ac} in-plane electric field. c) Spin textures for Rashba spin-orbit coupling in the absence of the magnetic field. (d) Energy spectrum of graphene with valley-Zeeman spin-orbit coupling for $\lambda_\mathrm{Z}<\mu$. The vertical arrow shows a transition induced by an in-plane {\em ac} magnetic field. An out-of-plane magnetic field does not induce any transitions in this case. (e) Same spectrum as in (d) with arrows showing transitions induced by an {\em ac} in-plane electric field.
(f) Spin structure of the ground state for valley-Zeeman spin-orbit coupling. The crosses (dots) represent spins polarized into (out of) the plane.}
 \end{figure}

The energy spectrum for a realistic case of $\lr<\mu$ is shown in Fig.~\ref{dispersionR}a, where we choose $\mu>0$ without  loss of generality. The chiral nature of the bands is evident from the expectation values of the spin operators in  each of the subbands:
\beq
\label{spinavg}
\langle\alpha\beta;\tau_z| \hat{\mathcal{S}}_{x}|\alpha\beta;\tau_z \rangle = - \frac{\beta\sin\theta}{\sqrt{1+r^2}},~~
\langle\alpha\beta;\tau_z| \hat{\mathcal{S}}_{y}|\alpha\beta;\tau_z\rangle = \frac{\beta\cos\theta}{\sqrt{1+r^2}},~~
\langle\alpha\beta;\tau_z| \hat{\mathcal{S}}_{z} |\alpha\beta;\tau_z\rangle= 0,
\eeq 
where $\hat{\mathcal{S}}_a\equiv \hat s_a\hat{\sigma}_0$ are the components of the spin operator and $r\equiv \lambda_{\rm R}/2v_Fk$. It is clear from Eq.~\er{spinavg} that $\bk\cdot\hat{\boldsymbol{\mathcal{S}}}=0$,
which means that the spin is perpendicular to the momentum in each subband, as shown in   Fig.~\ref{dispersionR}c. Note that $\langle\alpha\beta;\tau_z| \hat{\mathcal{S}}_{i}|\alpha\beta;\tau_z \rangle$ is independent of $\tau_z$. This means that the chiral structure is the same at the $K$ and $K'$ points.

To determine which transitions can be excited by an {\em ac} magnetic field in an ESR measurement, we also need the intravalley matrix elements of the spin operator for off-diagonal transitions between spin-split subbands. Using the eigenstates from Eq.~\er{es}, we obtain
\bse
\bea
\langle\alpha'\beta';\tau_z|\hat{\mathcal{S}}_x|\alpha\beta;\tau_z\rangle&=&\frac{i\left(\alpha'\beta' e^{i\theta}-\alpha\beta e^{-i\theta}\right)\left(\epsilon_{\alpha\beta}+\epsilon_{\alpha'\beta'}\right)}{2\sqrt{\left(1+\epsilon^2_{\alpha\beta}\right)\left(1+\epsilon^2_{\alpha'\beta'}\right)}},\label{sx}\\
\langle\alpha'\beta';\tau_z|\hat{\mathcal{S}}_y|\alpha\beta;\tau_z\rangle&=&\frac{\left(\alpha'\beta' e^{i\theta}+\alpha\beta e^{-i\theta}\right)\left(\epsilon_{\alpha\beta}+\epsilon_{\alpha'\beta'}\right)}{2\sqrt{\left(1+\epsilon^2_{\alpha\beta}\right)\left(1+\epsilon^2_{\alpha'\beta'}\right)}},\label{sy}\\
\langle\alpha'\beta';\tau_z|\hat{\mathcal{S}}_z|\alpha\beta;\tau_z\rangle&=&\frac{\left(\alpha\beta\alpha'\beta'-1\right)\left(1+\epsilon_{\alpha\beta}\epsilon_{\alpha'\beta'}\right)}{2\sqrt{\left(1+\epsilon^2_{\alpha\beta}\right)\left(1+\epsilon^2_{\alpha'\beta'}\right)}}.\label{sz}
\eea 
\ese
A response to an in-plane {\em ac} magnetic field is controlled by the matrix elements in Eqs.~\eqref{sx} and \eqref{sy}.  These vanish if $\epsilon_{\alpha\beta}+\epsilon_{\alpha'\beta'}=0$, which can happen only if $\beta=\beta'$ and $\alpha=-\alpha'$. Thus  transitions between conduction and valence subbands of the same chirality are forbidden. From Fig.~\ref{dispersionR}a, we see that the frequencies of the allowed transitions for an in-plane field are $\Omega=\lr$ and $\Omega=2\mu\pm\lr$. 
A response to an out-of-plane magnetic field is controlled by the matrix element in Eq.~\er{sz}, which is non-zero if  $\alpha\beta\alpha'\beta'\neq1$. This condition allows for transitions between the states with opposite subband (conduction vs valence) or chirality indices but not both. As shown in Fig.~\ref{dispersionR}a, the frequencies of the allowed transitions for an out-of-plane field are  $\Omega=\lr$ and $\Omega=2\mu$.

\begin{figure}
\centering
\includegraphics[scale=0.5]{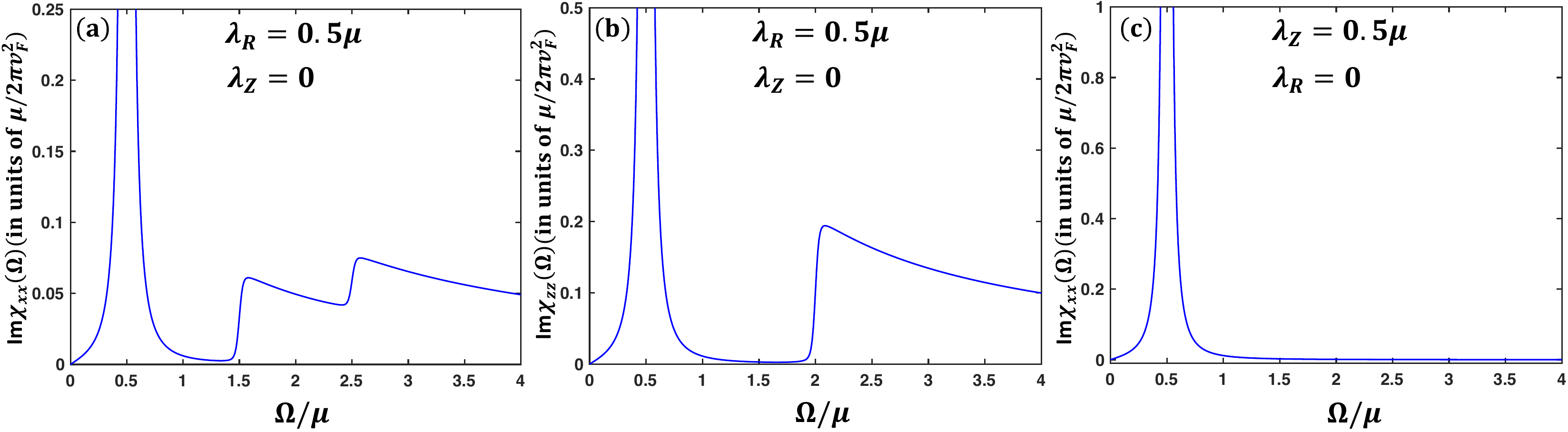}
\caption{\label{NonInt_Spin} {\em Non-interacting electrons}. Imaginary parts of the in-plane (a) and out-of-plane (b) spin susceptibilities for the case of Rashba SOC only, showing the resonance at $\lr$ and onsets of continua of conduction-to-valence band transitions at $\Omega=2\mu\pm\lr$ and  $\Omega=2\mu$. For the valley-Zeeman SOC case only (c), the out-of-plane susceptibility is zero. There is a resonance at $\lz$ in the in-plane susceptibility, but no continua. To mimic the effect of spin-relaxation processes, the electron states were broadened by $\gamma=0.005\mu$.}
\end{figure}

In an ESR experiment, one measures the imaginary part of the spin susceptibility, which can be calculated using the Kubo formula, see Appendix \ref{app:free_chi}. The results are shown in Figs.~\ref{NonInt_Spin}a and \ref{NonInt_Spin}b. In agreement with the selection rules, there is a resonance at $\Omega=\lr$ both in $\im\chi_{xx}(\Omega)$ and $\im\chi_{zz}(\Omega)$, and onsets of continua at $\Omega=2\mu\pm \lr$ in $\im\chi_{xx}(\Omega)$ and at  $\Omega=2\mu$ in $\im\chi_{zz}(\Omega)$.

\begin{figure}
\centering
\includegraphics[scale=0.45]{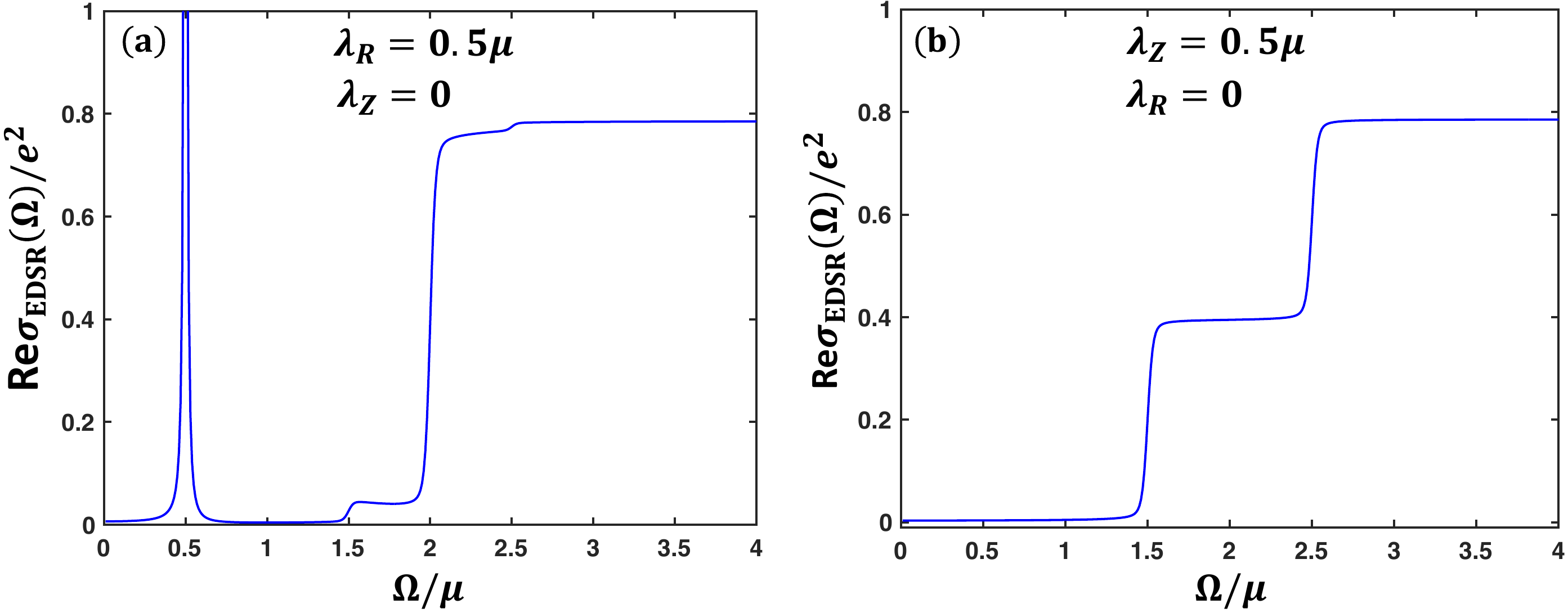}
\caption{\label{NonInt_Cond}{\em Non-interacting electrons}. (a) Real part of optical conductivity for Rashba SOC, showing the resonance at $\lr$ and continua starting at $\Omega=2\mu$ and $\Omega=2\mu\pm\lr$. The resonance is the zero-field EDSR effect. For the VZ case (b), there are continua starting at $\Omega=2\mu\pm\lz$, but no resonance at $\lz$. VZ SOC does not lead to zero-field EDSR resonance. In both cases, the Drude part of the conductivity is not shown. Broadening is the same as in Fig.~\ref{NonInt_Spin}.}
\end{figure}

To understand the selection rules for transitions induced by an in-plane {\em ac} electric field in an EDSR measurement, we need the matrix elements of the velocity operator, which coincides with the velocity operator for Dirac fermions without SOC coupling:
\beq
\label{vel}
\hat{\bv}=\boldsymbol{\nabla}_\bk \hat H_0= v_F\tau_z 
\hat s_0\hat\sigma_x\hat x +v_F\hat s_0\hat\sigma_y
\hat y.
\eeq
Using the eigenstates from Eq.~\er{es}, we obtain
\bea
\label{vmatR}
\langle\alpha' \beta';\tau_z| \hat v_x
|\alpha \beta;\tau_z \rangle &=&\tau_zv_F \frac{\epsilon_{\alpha\beta} e^{-i\theta\tau_z}+\epsilon_{\alpha'\beta'} e^{i\theta\tau_z} +\alpha\alpha'\beta\beta'\left(\epsilon_{\alpha\beta}e^{i\theta\tau_z}+\epsilon_{\alpha'\beta'}e^{-i\theta\tau_z}\right)}{2\sqrt{1+\epsilon_{\alpha\beta}^2}\sqrt{1+\epsilon_{\alpha'\beta'}^2}},\nn\\
\langle \alpha' \beta';\tau_z|\hat
v_y
|\alpha \beta;\tau_z \rangle &=& i\tau_zv_F \frac{\epsilon_{\alpha\beta} e^{-i\theta\tau_z}-\epsilon_{\alpha'\beta'} e^{i\theta\tau_z} -\alpha\alpha'\beta\beta'\left(\epsilon_{\alpha\beta}e^{i\theta\tau_z}-\epsilon_{\alpha'\beta'}e^{-i\theta\tau_z}\right)}{2\sqrt{1+\epsilon_{\alpha\beta}^2}\sqrt{1+\epsilon_{\alpha'\beta'}^2}}.
\eea
In contrast to the case of magnetic driving, we see that the matrix elements of the velocity are finite for any combination of $\alpha$, $\beta$, $\alpha'$, and $\beta'$, and thus transitions between all the subbands can be excited by an {\em ac} electric field (except for those within the valence bands, which are forbidden by the Pauli principle). Therefore, we expect to see a resonance in the conductivity at $\Omega=\lr$ and onsets of continua at $\Omega=2\mu\pm \lr$ and $\Omega=2\mu$, as shown in Fig.~\ref{dispersionR}b. In an EDSR experiment, one measures the real part of the optical conductivity, which can be also calculated using  the Kubo formula (see. Appendix \ref{app:free_sigma}). The conductivity shown in Fig.~\ref{NonInt_Cond}a (without the Drude part) indeed exhibits all the features following from the selection rules.

\subsection{Energy spectrum and selection rules for valley-Zeeman spin-orbit coupling}
\label{sec:free_VZ}
In the opposite limiting case, when only VZ SOC is present ($\lr=0$), the eigenvalues and eigenstates of the Hamlitonian \er{ham1} are given by \beq
\label{es_vz}
\begin{split}
\ve_{\alpha \beta;\tau_z}(\bk) &= \alpha\left(v_Fk + \beta \frac{\lambda_{\rm Z}}{2}\right),\\
\ket{\alpha \beta;\tau_z} &= \frac{1}{2\sqrt{2}} 
\left( \begin{array}{c}
1 + \tau_z \alpha \beta \\
(\tau_z \alpha + \beta)e^{i \tau_z \theta} \\
1 - \tau_z \alpha \beta  \\ 
(\tau_z \alpha - \beta)e^{i \tau_z \theta}
\end{array} \right)
\end{split}
\eeq
with the same  notations as in Eq.~\eqref{es}. 
The corresponding spectrum is shown in Fig.~\ref{dispersionR}b.

The expectation values of the spin operators are
\beq
\label{spinavgvz}
\langle\alpha\beta;\tau_z|\hat{\mathcal{S}}_{x}|\alpha\beta;\tau_z \rangle = 0,~~
\langle\alpha\beta;\tau_z|\hat{\mathcal{S}}_{y}|\alpha\beta;\tau_z \rangle = 0,~~
\langle\alpha\beta;\tau_z|\hat{\mathcal{S}}_{z}|\alpha\beta;\tau_z \rangle = \tau_z\alpha\beta.
\eeq 
In contrast to the case of Rashba SOC, spins are no longer chiral but Ising-like, and are polarized in the opposite directions in the SOC-split bands, see Fig.~\ref{dispersionR}d. Within a given electron or hole spin-split subband,  the direction of polarization is also opposite at the $K$ and $K'$ points. This is expected as the system preserves time-reversal symmetry. 

To understand the selection rules for transitions probed by ESR, we need the off-diagonal matrix elements of the spin operator, which are given by
\beq
\label{vzoff}
\langle\alpha'\beta';\tau_z|\hat{\mathcal{S}}_{x}|\alpha\beta;\tau_z \rangle = - \delta_{\alpha\alpha'} \delta_{\beta,-\beta'},~~
\langle\alpha'\beta';\tau_z|\hat{\mathcal{S}}_{y}|\alpha\beta;\tau_z \rangle =i\tau_z\alpha\beta \delta_{\alpha\alpha'} \delta_{\beta,-\beta'},~~
\langle\alpha'\beta';\tau_z|\hat{\mathcal{S}}_{z}|\alpha\beta;\tau_z \rangle = \tau_z\alpha\beta \delta_{\alpha\alpha'} \delta_{\beta\beta'}.
\eeq 
That $\langle\alpha'\beta';\tau_z|\hat{\mathcal{S}}_{z}|\alpha\beta;\tau_z \rangle$ is purely diagonal implies that an out-of-plane magnetic field cannot induce any inter-band transitions; this is so because $\hat{\mathcal{S}}_{z}$ commutes with the Hamiltonian. Furthermore, an in-plane magnetic field can only induce a transition at $\Omega=\lambda_\mathrm{Z}$ between the spin-split branches of the conduction band, but there are no continua of spin-flip transitions, see Fig.~\ref{dispersionR}d. Correspondingly, $\im\chi_{zz}(\Omega)=0$ while  $\im\chi_{xx}(\Omega)$ exhibits a single peak at $\Omega=\lz$, as shown in Fig.~\ref{NonInt_Spin}c.

The selections rules for transitions probed by an {\em ac} electric field follow from the matrix elements of velocity, given by \bea
\label{vmatZ}
\langle\alpha' \beta';\tau_z| \hat v_x
|\alpha \beta;\tau_z \rangle &=& \frac {\tau_zv_F}{4}\left[\alpha e^{i\theta\tau_z}\alpha'+e^{-i\theta\tau_z}\right](1+\alpha\alpha'\beta\beta'),\nn\\
\langle\alpha' \beta';\tau_z| \hat v_y
|\alpha \beta;\tau_z \rangle &=&  \frac{\tau_zv_F}{4i}\left[\alpha e^{i\theta\tau_z}-\alpha'e^{-i\theta\tau_z}\right](1+\alpha\alpha'\beta\beta').
\eea
It follows from the last equation that an in-plane {\em ac} electric field  can induce transitions  only between the valence and conduction band, with a simultaneous flip of chirality. As shown in Fig.~\ref{dispersionR}e, the frequencies of these transitions are $\Omega=2\mu\pm \lz$, but there is no resonance peak at $\Omega=\lz$. Indeed, the corresponding conductivity in Fig.~\ref{NonInt_Cond}b shows no resonance but only a double step at $\Omega=2\mu\pm \lz$, which is just a Pauli threshold at $\Omega=2\mu$ split by VZ SOC.

\subsection{Low-energy Hamiltonian for the conduction band in the presence of external electric and magnetic fields}\label{Sec:spinX}
As we saw in the previous sections, there are two groups of transitions that can be excited by external electric and magnetic fields: i) those between the spin-split subbands of the conduction band at $\Omega=\lr,\lz$ and ii) those between the spin-split conduction and valence bands at $\Omega=2\mu,2\mu\pm \lr,2\mu\pm\lz$. Given that $\lr,\lz\ll \mu$ in real systems,  the transition frequencies in the second group are near the direct absorption threshold at $\Omega\approx 2\mu$.  If the Coulomb interaction between
optically excited holes and conduction electrons is accounted for,  absorption starts at the indirect threshold of $\Omega=\mu$, via the same mechanism as in Auger damping of photoexcited carriers in doped semiconductors.\cite{Gavoret:1969,Pimenov:2017,Goyal} In addition, the interaction  between electrons in the conduction band also leads to absorption for $\Omega<2\mu$,\cite{Sharma:2021} and this contribution is comparable to Auger's one for $\Omega\sim \mu$. For $\Omega\gtrsim \mu$, the linewidth due to both types of damping is large, on the order of $g^2\mu$, where $g$ is the dimensionless coupling constant of the Coulomb interaction. Therefore, for moderate and strong interaction, both Auger and intraband damping are expected to smear the fine features near $2\mu$, induced by SOC.  For this reason, we will ignore transitions at $\Omega\approx 2\mu$ and focus on the low-energy part of the spectrum at $\Omega\approx \lr,\lz\ll\mu$.

For this range of frequencies, it makes sense to derive an effective low-energy Hamiltonian for the spin-split conduction band. We start by transforming Eq.~\er{ham1}, written in the sub-lattice basis $\{\hat a_\uparrow,\hat b_\uparrow,\hat a_\downarrow,\hat b_\downarrow\}^T$,  to the subband basis  $\{\hat c_\uparrow,\hat c_\downarrow,\hat v_\uparrow,\hat v_\downarrow\}^T$, in which the Dirac part of $\hat H_0$ is diagonal. The transformation is effected via
\bea
\label{basis}
\hat a_\vs=\frac{\hat c_\vs+\hat v_\vs}{\sqrt{2}},\;b_\vs=\frac{\hat c_\vs-\hat v_\vs}{\sqrt{2}}\tau_ze^{i\tau_z\theta},\;\vs={\uparrow,\downarrow}.\label{abcv}
\eea
An {\emph ac}  magnetic field $\hat{\bf{b}}_0B_0e^{-i\Omega t}$ is accounted for by adding the Zeeman term $\hat {\bf s}\cdot \hat {\bf b}_0\hat\sigma_0 \Delta_{\mathrm{Z}}/2$ to the Hamiltonian \er{ham1}, where $\Delta_{\mathrm{Z}}=g\mu_B B_0e^{-i\Omega t}$, $g$ is the effective Land{\'e}-factor, and $\mu_B$ is the Bohr magneton.  An {\em ac} electric field ${\bf E}_0e^{-i\Omega t}$ is accounted for by a gauge transformation $\bk\to \bk+ (e/c){\bf A}$ in  Eq.~\er{ham1}, where ${\bf A}=(c/i\Omega){\bf E}_0 e^{-i\Omega t}$ is the vector-potential. In the new basis, and in the presence of both magnetic and electric fields, the Hamiltonian can be written as a $4\times 4$ block-matrix  \bea
\hat H_0^{\rm band}=\left(
\begin{array}{cc}
\hat H_{cc} & \hat H_{cv}\\
\hat H_{vc} & H_{vv}
\end{array}
\right),
\label{Hcv}
\eea
where the $2\times 2$ blocks are given by
\bea
\label{HR_cc}
\hat H_{cc}&=&\hat s_0 v_F k + \frac{\lr}{2} (\hat\bk\times \hat{\bf s})\cdot\hat{\bf z}+\frac{\lz}{2}\tau_z\hat s_z+\frac{\Delta_{\mathrm{Z}}}{2} \hat{\bf{s}}\cdot \hat{\bf{b}}_0 
+\hat s_0\frac{ev_F}{c} (\hat\bk\cdot {\bf A}),
\nn\\
\hat H_{vv}&=&-\hat s_0 v_F k- \frac{\lr}{2} (\hat\bk\times \hat{\bf s})\cdot\hat{\bf z}+ \frac{\lz}{2}\tau_z\hat s_z
+
\frac{\Delta_{\mathrm{Z}}}{2} \hat{\bf{s}}\cdot \hat{\bf{b}}_0 
-\hat s_0\frac{ev_F}{c} (\hat\bk\cdot {\bf A}),\nn\\
\hat H_{cv}&=&i\tau_z\hat s_0\frac{ev_F}{c} (\hat\bk\times{\bf A})\cdot\hat{\bf z}-i \frac{\lr}{2}\tau_z(\hat\bk\cdot\hat{\bf s})\nn\\
\hat H_{vc}&=&-\hat H_{cv},
\eea
with $\hat\bk=\bk/k$ and $\hat{\bf z}$ being the unit vector normal the plane. Note that the coupling between the conduction and valence bands arises via the electric field and Rashba SOC. This is due to the selection rules discussed in Secs.~\ref{sec:free_RSOC} and \ref{sec:free_VZ}, which say that, at $\Omega\ll \mu$,  only Rashba SOC allows for transitions between the conduction and valence bands. Note also that the Rashba terms  in $\hat H_{cc}$ and $\hat H_{vv}$ are almost the same as in a 2D electron gas  (up to the dependence on the magnitude of $\bk$).

Next, we project out the valence band via a standard downfolding procedure. Namely, we find the Green's function of $\hat H_0^{\rm band}$ in Eq.~\er{Hcv}, $G=(\epsilon-\hat H_0^{\rm band})^{-1}$, and read off its $cc$ element. This yields the effective low-energy Hamiltonian as $\hat {\mathcal H}^{\rm proj}_{cc}=\hat H_{cc}+\hat H_{cv}(\epsilon-\hat H_{vv})^{-1} \hat H_{vc}$. To  leading order in the external fields and SOC, the eigenvalue $\epsilon$ can be replaced by $v_Fk$, leading to
\beq
\hat {\mathcal H}^{\rm proj}_{cc}=
\hat s_0 v_F k + \frac{\lr}{2} (\hat\bk\times \hat{\bf s})\cdot\hat{\bf z}+\frac{\lz}{2}\tau_z\hat s_z+\frac{\Delta_{\mathrm{Z}}}{2}\hat{\bf{s}}\cdot\hat{\bf{b}}_0
+\hat s_0\frac{ev_F}{c} (\hat\bk\cdot {\bf A})+\frac{e\lr}{2ck}
({\bf A}\times \hat\bk)\cdot\hat{\bf z}
(\hat\bk\cdot \hat{\bf s}).
\label{leR}
\eeq
Note that the electric field couples to spins only due to Rashba SOC, which is the origin of the EDSR effect. In the absence of external fields, SOC splits the conduction band into two subbands with energies 
\beq\label{spec_both}\ve_{\pm}=v_Fk\pm\frac 12
\lambda_{\text{SOC}},\;\text{where}\;\lambda_{\text{SOC}}=\sqrt{\lr^2+\lz^2},
\eeq 
and the resonance  occurs at $\Omega=\lambda_{\text{SOC}}$.

\section{Electron spin and electric-dipole spin resonances in a two-valley Fermi liquid}\label{Sec:IntModel}
\subsection{Two-valley Fermi liquid}
\label{sec:2VFL}
In this section, we investigate the effect of electron-electron interaction on ESR and EDSR.  As we shall demonstrate, the most prominent effect of the interaction is to split the resonances into two. This splitting is controlled by the coupling constants of the various interaction channels, which enables one to extract these important parameters from the measured spectra. Since we are interested only in energies much smaller than the Fermi energy, the effect of electron-electron interaction can be accounted for within a Fermi liquid (FL) theory for conduction electrons only, while the interaction with holes can be assumed to be absorbed into the coupling constants of the FL theory. 

\begin{figure}
\centering
\includegraphics[scale=0.45]{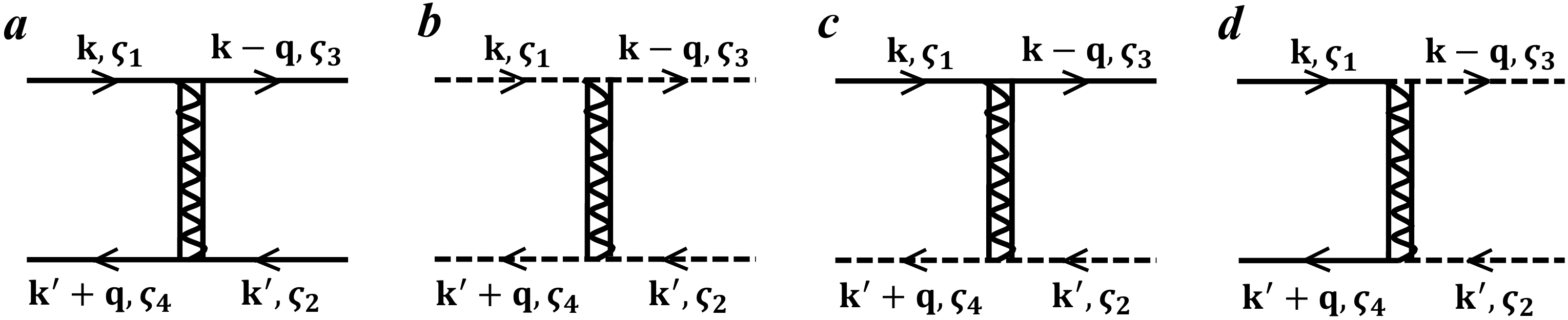}
\caption{\label{scatt} Interaction vertices for intra-valley ($a$ and $b$) and inter-valley ($c$ and $d$) scattering processes. The solid (dashed) lines refer to electrons in the $K$ ($K'$) valley. Diagrams $a$ and $b$ also have exchange partners with outgoing states swapped (not shown). Diagram $d$ involves a  large momentum transfer $\sim|\bold{K}-\bold{K}'|$, and is neglected in our model.}
\end{figure}

 First, we discuss the structure of the FL theory for a two-valley system in the absence of SOC, developed recently in Ref.~\onlinecite{Raines:2021}. The interaction  vertices are shown in Fig.~\ref{scatt}.
The solid and dashed lines depict electrons in the $K$ and $K'$ valleys, respectively. 
Diagrams $a$ and $b$ describe intra-valley scattering. Each of these diagrams has an exchange partner  (not shown), in which the outgoing states are swapped. 
Diagram $c$ describes an inter-valley scattering event, in which electrons stay in their respective valleys. The momentum transfer in such an event is less than or equal to $2k_F$. Diagram $d$ is an exchange partner to diagram $c$, in which electrons are swapped between the valleys. The momentum transfer in such an event is close to the distance between the $K$ and $K'$ points, $|{\bf K}-{\bf K'}|\sim 1/a$, where $a$ is the lattice constant. For $k_Fa\ll 1$, the matrix element of the Coulomb interaction for diagram $d$ is much smaller than that for diagram $c$, and will be neglected in what follows. In this case, the valley index plays a role of conserved isospin, and we have an $SU(2)\times SU(2)$-invariant FL. The interaction between quasiparticles of such a FL  is described by the Landau interaction function \bea
\label{FL}
\nu_F^* \hat{f}(\bk,\bk') &=& F^s(\bk,\bk') \delta_{\upsilon_1 \upsilon_3} \delta_{\upsilon_2 \upsilon_4} \delta_{\varsigma_1 \vs_3} \delta_{\vs_2 \vs_4} + F^a(\bk,\bk') \delta_{\upsilon_1 \upsilon_3} \delta_{\upsilon_2 \upsilon_4} (\hat{\bf{s}}_{\vs_1 \vs_3} \cdot \hat{\bf{s}}_{\vs_2\vs_4})\nn\\
&+&G^a(\bk,\bk') (\btau_{\upsilon_1 \upsilon_3} \cdot \btau_{\upsilon_2 \upsilon_4}) \delta_{\vs_1 \vs_3} \delta_{\vs_2 \vs_4}+ H(\bk,\bk') (\btau_{\upsilon_1 \upsilon_3} \cdot \btau_{\upsilon_2 \upsilon_4}) (\hat{\bf{s}}_{\vs_1 \vs_3} \cdot \hat{\bf{s}}_{\vs_2 \vs_4}),
\eea
where $\nu_F^*$ is the renormalized density of states at the Fermi energy and $v$ ($\vs$) labels valley (spin). Components $F^s$ and $F^a$, which are present also in the single-valley case, describe direct and exchange interaction between fermions in the same valley, respectively. Component $G^a$ describes exchange interaction between different valleys, which would be present even for spinless fermions. Finally, component  $H$ describes exchange interaction between both spins and valleys.
The components of the Landau function are defined on the Fermi surface, i.e., for $k=k'=k_F$, depend only on angle $\vartheta$ between $\bk$ and $\bk'$, and can be characterized by angular harmonics, e.g., 
$F^a_m=\int d\vartheta F^a(\vartheta) e^{im\vartheta}/2\pi$, and the same for other components.

As long as SOC can be treated as a perturbation, i.e., for $\lr,\lz\ll\mu$, the Landau function in Eq.~\er{FL} can also be used to describe dynamics of spins in the presence of SOC, similar to how it was done in Refs.~\onlinecite{shekhter2005, ashrafi2013, kumar2017} for a 2D electron gas with Rashba and Dresselhaus SOC. 

\subsection{Collective modes of a two-valley Fermi liquid with spin-orbit coupling}
\label{sec:eigen}

Accounting for the valley degree of freedom and in the absence of external fields, one can write the projected Hamiltonian \er{leR} becomes
\beq
\hat{\mathcal{H}}^{\rm proj}_{cc} = v_F k \hat{\tau}_0 \hat s_0 + \frac{\lr}{2} \hat{\tau}_0 (\hat\bk\times \hat{\bf s})\cdot\hat{\bf z} + \frac{\lambda_Z}{2} \hat{\tau}_z \hat{s}_z.
\eeq
Accordingly, the quasiparticle energy can be written as the sum of the equilibrium part (eq) and a correction due to the Landau functional (LF), whereas the equilibrium part is further separated into a spin-independent part and a correction due to SOC:
\bse
\bea
\label{sce}
\hat{\ve}(\bk,t) &=&\hat\ve_{\text{eq}}(\bk)+\delta\hat\ve_{\mathrm{LF}}(\bk,t),\\
\hat\ve_{\text{eq}}(\bk)&=&\hat\tau_0 \hat s_0v^*_F(k-k_F)+\delta \hat{\ve}_{\text{SO}}(\bk),\\
\delta \hat{\ve}_{\text{SO}}(\bk)&=&\frac{\lr^*}{2} \hat{\tau}_0
(\hat\bk\times \hat{\bf s})\cdot\hat{\bf z}
 + \frac{\lz^*}{2}\hat{\tau}_z \hat{s}_z,\label{eso}\\
\delta\hat\ve_{\mathrm{LF}}(\bk,t)&=&\text{Tr}' \int \frac{d^2p'}{(2\pi)^2} \hat{f}(\bk, \bk')\hat{n}(\bk',t).\label{sce_d}
\eea
\ese
Here, $\delta \hat{n}(\bk,t)$ is the density matrix, $f(\bk,\bk')$ is given by Eq.~\er{FL}, $'$ indicates the valley/spin state of a quasiparticle with momentum $\bk'$, $v_F^*$ is the renormalized Fermi velocity, and
$\lr^*=\lr/(1+F_1^a)$,\cite{shekhter2005} $\lz^*=\lz/(1+H_0)$\cite{Raines:2021b}  are the renormalized spin-orbit coupling constants. Equations \er{sce}-\er{sce_d} need to be solved self-consistently along  with the kinetic equation for the density matrix
\beq
\label{kinetic}
i \frac{\partial \hat{n} (\bk, t)}{\partial t} = [ \hat{\ve} (\bk, t), 
\hat{n} (\bk, t)].
\eeq
[Our kinetic equation does not contain the effects of Berry curvature because we neglected the asymmetry gap in the single-particle spectrum.]
As in Refs.~\onlinecite{shekhter2005, kumar2017},
 we introduce a set of rotated Pauli matrices in the spin space
\beq
\label{rot}
\hat\xi_0=\hat s_0, \,\,\,\,\,\, \hat{\xi}_1(\bk) = -\hat{s}_z, \,\,\,\,\,\, \hat{\xi}_2(\bk) = \cos\theta \hat{s}_x + \sin\theta \hat{s}_y, \,\,\,\,\,\, \hat{\xi}_3(\bk) = \sin\theta \hat{s}_x - \cos\theta \hat{s}_y,
\eeq
and parametrize $\delta\hat{n}({\bf p,t})$ as the sum of the equilibrium (eq) and fluctuating (fl) parts: 
\cite{Raines:2021,Raines:2021b}
\bse
\bea
\hat n(\bk,t)&=&\hat n_{\text{eq}}(\bk)+\delta\hat n_{\text{fl}}(\bk,t),\label{dn_a}\\
n_{\text{eq}}(\bk)&=&\hat\tau_0\hat\xi_0 n_F+n_F'\hat\ve_{\text{SO}}(\bk),\\
\delta \hat{n}_{\text{fl}}(\bk, t)& =& n_F' \hat\tau_0\hat\xi_0 a(\bk,t)+\delta \hat n_{\text{sv}}(\bk,t),\label{dn} \\
 \delta {\hat n_\text{sv}}(\bk,t)&=&n_F'\left[\hat{\tau}_0 \bu(\bk, t) \cdot \bxi(\bk) + \bw(\bk, t) \cdot \btau \hat{\xi}_0 + M_{\alpha\beta}(\bk, t) \hat{\tau}_\alpha \hat{\xi}_\beta(\bk)\right]\label{dnsv},
\eea
\ese
{where $n_F$ is the Fermi function, $\alpha,\beta \in \{1,2,3\}$, and $n_F'\equiv \partial_\ve n_F(\ve)$. Vector $\bu$ and tensor $M_{\alpha\beta}$  describe oscillations of the uniform magnetization
\beq
\tilde S_\alpha=-\frac{g\mu_B}{2}\mathcal{S}_{\alpha}=-\frac{g\mu_B}{2} \int \frac{d^{2}k}{(2\pi)^2}  \text{Tr}\left[ \delta \hat n(\bk,t)\, \hat s_\alpha\right]=\frac{g\mu_B\nu_F^*}{8} \int \frac{d\theta_\bk}{2\pi}  u_\beta (\bk, t) \text{Tr}\left[\hat \xi_\beta \hat s_\alpha\right]\label{magn}
\eeq
and valley-staggered magnetization
\bea
\tilde M_\alpha=-\frac{g\mu_B}{2}\mathcal{M}_\alpha=-\frac{g\mu_B}{2} \int \frac{d^{2}k}{(2\pi)^2}  \text{Tr}\left[ \delta \hat n(\bk,t)\,\tau_z \hat s_\alpha\right]=\frac{g\mu_B\nu_F^*}{8} \int \frac{d\theta_\bk}{2\pi} \mathcal{N}_{\beta}(\bk,t)\text{Tr}\left[\hat\xi_\beta \hat s_\alpha\right],\label{vsmagn}
\eea
respectively, where $\mathcal{N}_{\gamma}(\bk,t)\equiv M_{3\gamma}(\bk,t)$ with $\gamma=1\dots 3$. Vector ${\bf w}$ describes oscillations in the valley occupancy, which are decoupled from both magnetizations and  will not be considered below. 
Expanding  $\bu$ and $\boldsymbol{\mathcal{N}}$ over the set of angular harmonics as $\bu(\bk,t)=\sum_m e^{im\theta} \bu^{(m)}(t)$ and $\boldsymbol{\mathcal{N}}(\bk,t)=\sum_m e^{im\theta} \boldsymbol{\mathcal{N}}^{(m)}(t)$, we obtain for the components of $\boldsymbol{\mathcal{S}}$ 
\bea
\mathcal{S}_x=\nu_F^*\left[ \frac{u_2^{(-1)}+u_2^{(+1)}}{2}+\frac{u_3^{(-1)}-u_3^{(+1)}}{2i} \right],\;
\mathcal{S}_y=\nu_F^*\left[ \frac{u_2^{(-1)}-u_2^{(+1)}}{2i}-\frac{u_3^{(-1)}+u_3^{(+1)}}{2} \right],\;\mathcal{S}_z=-\nu_F^*u_1^{(0)}.\label{sxyz}
\eea
The expressions for $\mathcal{M}_i$ are obtained from the last equation by replacing $u_{\gamma}^{(\pm 1)}\to \mathcal{N}_{\gamma}^{(\pm 1)}$. 

The equations of motion for $u^{(m)}_\alpha(t)$ and $\mathcal{N}^{(m)}_\alpha(t)$ are obtained by tracing out the corresponding components of Eq.~\er{kinetic}. The full system of equations is presented in Appendix~\er{Kin}. To understand the dynamics of the system, it is instructive to consider first the case of $\lambda_Z^*=0$, when the equations for $u^{(m)}_\alpha(t)$ and $\mathcal{N}^{(m)}_\alpha(t)$  decouple. The $m=\pm 1$ harmonics, which enter the in-plane components of $\boldsymbol{\mathcal{S}}$ and $\boldsymbol{\mathcal{M}}$, satisfy
\bse
\bea
\dot{u}_1^{(\pm 1)} &=& \lr^* \left[f_+ u_2^{(\pm 1)} \pm i f_- u_3^{(\pm 1)} \right],\;
\dot{u}_2^{(\pm 1)} = -\lr^* f u_1^{(\pm 1)},\dot{u}_3^{(\pm 1)} =0;\label{equ}\\
\dot{\mathcal{N}}_1^{(\pm 1)} &=& \lr^* \left[h_+ \mathcal{N}_2^{(\pm 1)} \pm i h_- \mathcal{N}_3^{(\pm 1)} \right],\;
\dot{\mathcal{N}}_2^{(\pm 1)} = -\lr^* h \mathcal{N}_1^{(\pm 1)},\;
\dot{\mathcal{N}}_3^{(\pm 1)} =0,\label{eqN}
\eea
\ese
where 
\bea 
f&=&1+F^a_1,\;f_+=1+(F^a_0+F^a_2)/2,\; f_-= (F^a_0-F^a_2)/2,\nn\\
 h&=& 1+H_1,\;h_+=1+(H_0+H_2)/2,\;  h_-= (H_0-H_2)/2. 
\label{fh}
\eea
Equations \er{equ} and \er{eqN} describe independent oscillations of  the in-plane magnetization and valley-staggered magnetization with frequencies
$\Omega_{-}=\lr^*\sqrt{ ff_+}$ and  $\Omega_{+}= \lr^* \sqrt{hh_+}$, respectively. From the structure of eigenvectors, one can deduce that both modes are linearly polarized.

The out-of-plane components of  $\boldsymbol{\mathcal{S}}$ and $\boldsymbol{\mathcal{M}}$ are expressed via the $m=0$ harmonics of $\bu$ and $\boldsymbol{\mathcal{N}}$, which satisfy  a set of equations similar to Eqs.~\er{equ} and \er{eqN}
with different combinations of the Landau parameters [see Appendix~\er{Kin}]. The mode frequencies in the $\boldsymbol{\mathcal{S}}$ and $\boldsymbol{\mathcal{M}}$ sectors are $\Omega_{-,z}= \lr^*\sqrt{f(f_{+}+f_{-})}$ and $\Omega_{+,z}= \lr^*\sqrt{h(h_{+}+h_{-})}$, respectively.

Valley-Zeeman SOC mixes up the $\boldsymbol{\mathcal{S}}$ and $\boldsymbol{\mathcal{M}}$ sectors. The frequencies of the in-plane modes (with $m=\pm1$) for $\lz^{*}\neq 0$ are given by
\bse
\beq
\label{R+Z}
\begin{split}
\Omega_\pm^2 =&
\lr^{*2}\bigg( \frac{ff_++hh_+}{2} \bigg) + \lz^{*2} \bigg( f_+h_++f_-h_- \bigg) 
\pm\Omega_{0}^2,
\end{split}
\eeq
where 
\beq
\label{O0}
\begin{split}
\Omega_{0}^2 =& \left[\lr^{*4}\bigg( \frac{ff_+-hh_+}{2}\bigg)^2
+\lz^{*4} (f_-h_++h_-f_+)^2+\lr^{*2} \lz^{*2}(ff_-+hh_-)(f_-h_++h_-f_+)\right]^{1/2}.
\end{split}
\eeq
\ese
In assigning the $\pm$ indices  to the modes, we assumed that $ff_+<hh_+$ and $f_{-}h_++h-_{-}f_+<0$. These two conditions are satisfied 
in the most realistic case, 
when i) both the intra- and intervalley interactions are attractive, i.e., $F^a_m,H_m<0$; ii) the intravalley interaction is stronger than the intervalley one, i.e., $|F^a_m|>|H_m|$; and iii) both $|F^a_{m}|$ and $|H_{m}|$ decrease with $m$ monotonically. In this case, $\Omega_+>\Omega_-$ for any ratio of $\lr^*$ to $\lz^*$. The structure of the eigenstates indicates that for both modes $\boldsymbol{\mathcal{S}}$ and $\boldsymbol{\mathcal{M}}$ are perpendicular to each other and phase lagged by $\pi/2$.

Solving the equations of motion for the $m=0$ harmonics of $u_\alpha$ and $\mathcal{N}_\alpha$, one finds that 
the out-of-plane modes oscillate with frequencies
\beq
\Omega_{+,z}^2=h\left[(h_++h_-)\lr^{*2}+f\lz^{*2}\right] \;\;\text{and} \;\; \Omega_{-,z}^2=f\left[(f_++f_-)\lr^{*2}+h\lz^{*2}\right],\eeq respectively.

One can  get a better insight into a general case by using explicit forms of the Landau  parameters, calculated to first order in the Hubbard interaction with amplitude $U_0$. \footnote{The range of the  interaction is assumed to be smaller than the Fermi wavelength but larger that the lattice constant, such that electrons are not transferred between the valleys.}
 The only non-zero Landau parameters in this approximation are\cite{Raines:2021}
$F^a_0=G^a_0=H_0=-u,$
and
$F^a_1=G^a_1=H_1=-u/2$,
where $u\equiv\nu_F U_0/8$. Note that, in contrast to a conventional FL, where a weak short-range interaction gives rise only to $m=0$ Landau parameters, a Dirac FL in graphene has at least the $m=0$ and $m=1$ harmonics. Correspondingly, $f=f_+=h=h_+=1-u/2
$ and $f_-=h_-=-u/2
$, and the in-plane mode frequencies are reduced to 
\bea
\label{Owc}
\Omega_\pm &=&\sqrt{\lr^{2}+\lz^{2}}
\left(1+\frac{u\lz^2}{2(\lr^2+\lz^2)}\right)
\pm
 \frac{u\lz}{2}.
\eea
A special feature of the weak Hubbard coupling is that the inter- and intravalley interactions are the same in this limit, and, therefore, the modes are degenerate at $\lz=0$. 
This degeneracy is lifted by the second term in Eq.~\er{Owc}, which is non-zero only if both VZ SOC and electron-electron interaction are present. 
\subsection{Zero-field electron spin resonance in a two-valley Fermi liquid}
\label{sec:esr}
To describe ESR, we assume that a weak, oscillatory magnetic field is applied  in the $y$-direction. Accordingly, the quasiparticle energy in Eq.~\eqref{sce}  acquires an extra term, $\delta \hat\ve_{B}(t) = (\Delta^*_Z/2) \hat{\tau}_0 \hat{s}_y$, where $\Delta_{\mathrm{Z}}^*=\Delta_{\mathrm{Z}}/(1+F_0^a)$ is the renormalized Zeeman energy.  The spin susceptibility is deduced from the  relation $\tilde{S}_i=\chi_{ij} B_j$, where $\tilde{S}_i$ is the $i^{\text{th}}$ component of the uniform magnetization, defined by Eq.~\er{magn}. Due to the rotational invariance of our model in the absence of the external field, the in-plane part of $\chi_{ij}$ is diagonal and symmetric. Solving the equations of motion for the density matrix (see Appendix \ref{Kin}), we find for the imaginary part of the in-plane spin susceptibility 
\bea
\text{Im}\chi(\Omega)=
\chi_0
\left[W^{\rm ESR}_+\Omega_+\delta(\Omega - \Omega_+)
+W^{\rm ESR}_-\Omega_{-}\delta(\Omega - \Omega_-) \right],\label{imchi}
\eea
where the resonance frequencies are given by Eqs.~\er{R+Z} and \er{O0},  $W^{\rm ESR}_{\pm}$ are the oscillator strengths, and $\chi_0
=g^2\mu_B^2\nu^*_F/4
(1+F_0^a)$
is the static spin susceptibility of a FL in doped graphene. The most interesting feature of the result in Eq.~\er{imchi} is that, in general, the ESR signal consists of two peaks rather than one, with weights
given by $W^{\rm ESR}_{\pm}=\pm W^{\rm ESR}(\Omega_\pm)$, where
\bea\label{WESR}
W^{\rm ESR}(\Omega)&=&\frac{\pi}{8}\frac{\Omega^2-\Omega^{2}_s}{\Omega^2 \Omega_0^2}[\lr^{*2}f+2\lz^{*2}(h_++h_-)],\;\text{and}\nn\\
\Omega_{s}^2 &=& \frac{[\lr^{*2}fh_++\lz^{*2}(h_+^2-h_-^2)][\lr^{*2}h+2\lz^{*2}(f_+-f_-)]}{\lr^{*2}f+2\lz^{*2}(h_++h_-)}.
\eea
\begin{figure}
\centering
\includegraphics[scale=0.5]{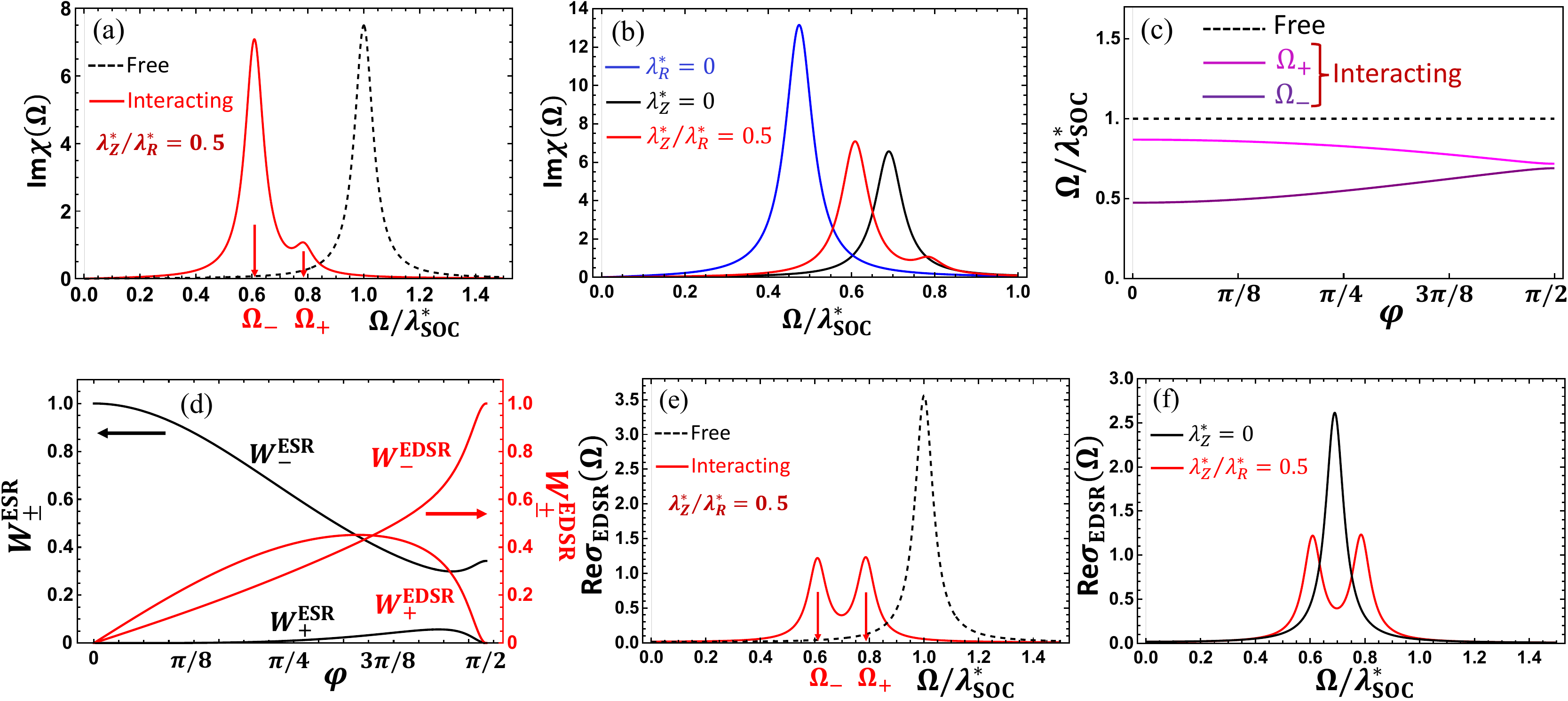}
\caption{\label{Peaks} Zero-field electron spin resonance (ESR) and electric-dipole spin resonance (EDSR) in graphene with proximity-induced spin-orbit coupling (SOC). (a) ESR signal. Vertical axis: the imaginary part of the dynamical spin susceptibility in units of $\chi_0$, defined in Eq.~\er{imchi}. The frequency on the horizontal axis is 
scaled with $\lambda^*_{\text{SOC}}=\sqrt{\lr^{*2}+\lz^{*2}}$, where $\lr^*$ and $\lz^*$ are (renormalized) couplings of the  Rashba and valley-Zeeman (VZ) types of SOC, respectively. $\Omega_{\pm}$ are the resonance frequencies, given by Eqs.~\er{R+Z} and \er{O0}. Dashed line: non-interacting system. Red solid line: a two-valley Fermi liquid (FL)  with parameters $F^a_0=-0.5500$, $F^a_1=-0.2750$, $F^a_2=-0.1375$,  $H_0=-0.5000$, $H_1=-0.2500$, and $H_2=-0.1250$.  The ratio $\lz^*/\lr^*=0.5$. The choice of FL parameters is the same for all panels of the figure.
 (b) ESR signal in a FL for several values of $\lz^*/\lr^*$, as indicated in the legend. (c) Resonance frequencies $\Omega_+$ and $\Omega_-$, given by Eqs.~\er{R+Z} and \er{O0},  as a function of angle $\varphi$, defined in Eq.~\er{angle}. $\varphi=0\,(\pi/2)$ corresponds $\lr^*=0$\, ($\lz^*=0$).
(d) Oscillator strengths of ESR (left vertical axis) and EDSR (right vertical axis) peaks as a function of angle $\varphi$, as given by Eqs.~\er{WESR} and \er{WEDSR}, respectively. 
(e)  EDSR signal. Vertical axis:  the real part of the optical conductivity in units of $\sigma_0$, defined in Eq.~\er{resigma}. Dashed line: non-interacting system. Solid line: FL. 
(f) EDSR signal in a FL for two values of $\lz^*/\lr^*$, as indicated in the legend.}
\end{figure}

The splitting of  the ESR signal occurs, first of all, because of electron-electron interaction. Indeed, for a non-interacting system, $f=f_+=h=h_+=1$ and $f_-=h_-=0$, such that the frequency $\Omega_0$ in Eq.~\er{O0} vanishes, and $\Omega_+=\Omega_-=\lambda_{\text{SOC}}$. This feature is demonstrated in Fig.~\ref{Peaks}a, where the dashed line depicts the ESR signal in the absence of interaction and the solid line depicts the same for a generic choice of the Landau parameters, as indicated in the figure caption. Splitting of the resonance occurs as long as the Landau function has more than just the $F_0^a$ harmonic, which, as was mentioned in Sec.~\ref{sec:eigen}, is always the case for graphene. In a real system, the widths of the resonances is controlled by spin-relaxation processes.  At low temperatures, the dominant mechanism of spin-relaxation is scattering by disorder in the presence of either extrinsic or intrinsic SOC. To account for this effect, we added a damping term, $-\gamma \delta\hat n(\bk,t)$, to the right-hand side of Eq.~\er{kinetic}. In all panels of  Fig.~\ref{Peaks}, $\gamma=0.04\lambda^*_{\text{SOC}}$, where $\lambda^*_{\text{SOC}}=\sqrt{\lr^{*2}+\lz^{*2}}$.  For $\lr^*=15.0$\,meV and $\lz^*=7.5$\,meV, the corresponding relaxation time $\gamma^{-1}=1$\,ps. 

However,  electron-electron interaction is a necessary but not sufficient condition for observing two ESR peaks. As shown in Fig.~\ref{Peaks}b, there is only one ESR peak, if only one of the two types of SOC is present. In this case,  the system still has two non-degenerate eigenmodes, but one of them is ESR-silent because the corresponding oscillator strength vanishes.   For example, if $\lz^*=0$ then, as we saw in Sec.~\ref{sec:eigen}, the two eigenmodes correspond to decoupled oscillations of the uniform and valley-staggered magnetizations.
But the frequency $\Omega_{+}$ of the valley-staggered mode for $\lz^*=0$ coincides with $\Omega_s$ in Eq.~\er{WESR}, and thus the corresponding oscillator strength, $W^{\text{ESR}}_-$ in \er{WESR}, vanishes, leaving the valley-staggered mode silent. Likewise, the $\Omega_+$ mode is also silent if $\lr^*=0$.

The dependence of the resonance frequencies on the ratio of the Rashba and VZ couplings is demonstrated in Fig.~\ref{Peaks}c. Here, the angle $\varphi\in[0,\pi/2]$ is defined as 
\bea
\label{angle}
\cos\varphi = \frac{\lz^{*}}{\sqrt{\lr^{*2}+ \lz^{*2} }}\equiv\frac{\lz^*}{\lambda^*_{\text{SOC}}},
\eea
such that $\varphi=0$ for $\lr^*=0$ and $\varphi=\pi/2$ for $\lz^*=0$.

The oscillator strengths of the two ESR peaks as a function of $\varphi$ are shown in Fig.~\ref{Peaks}d by the solid black lines.
As we see from panels (a) and (d),  the valley-staggered mode is much weaker than the uniform-magnetization one: for a particular choice of the Landau parameters, the oscillator strength of the former is about 20\% of the latter. The ratio of the two oscillator strengths is quite sensitive to the choice of  Landau parameters, in particular, to the comparative strength of the interaction in the spin  ($F^a_m$) and spin-valley ($H_m$) sectors. In our case, the Landau parameters form a 6-dimensional space. To restrict the parameter space,  we choose $F_0^a = 2F_1^a = 4F_2^a$, $H_0 = 2H_1 = 4H_2$, and $H_0 = a F_0^a$ with $0<a<1$, and plot the ratio  $W^{\text{ESR}}_{+}/W^{\text{ESR}}_{-}$ as a function of angle $\varphi$ at fixed $F_0^a$ and for several values of $a$; cf. Fig.~\ref{residue}a.  As we see from the plot, the maximum value of the ratio starts from 5\% for $a=0$ but increases towards 100\% as  $a$ approaches $1$.  It needs to be kept in mind though that the 100\% ratio is achieved only at $\varphi=\pi/2$, i.e., at $\lz^*=0$, when the two modes become degenerate. (For $a>1$, the  modes are swapped:  the frequency of the uniform mode becomes $\Omega_{+}$ while the frequency of the valley-staggered mode becomes $\Omega_{-}$.)

\begin{figure}
\centering
\includegraphics[scale=0.5]{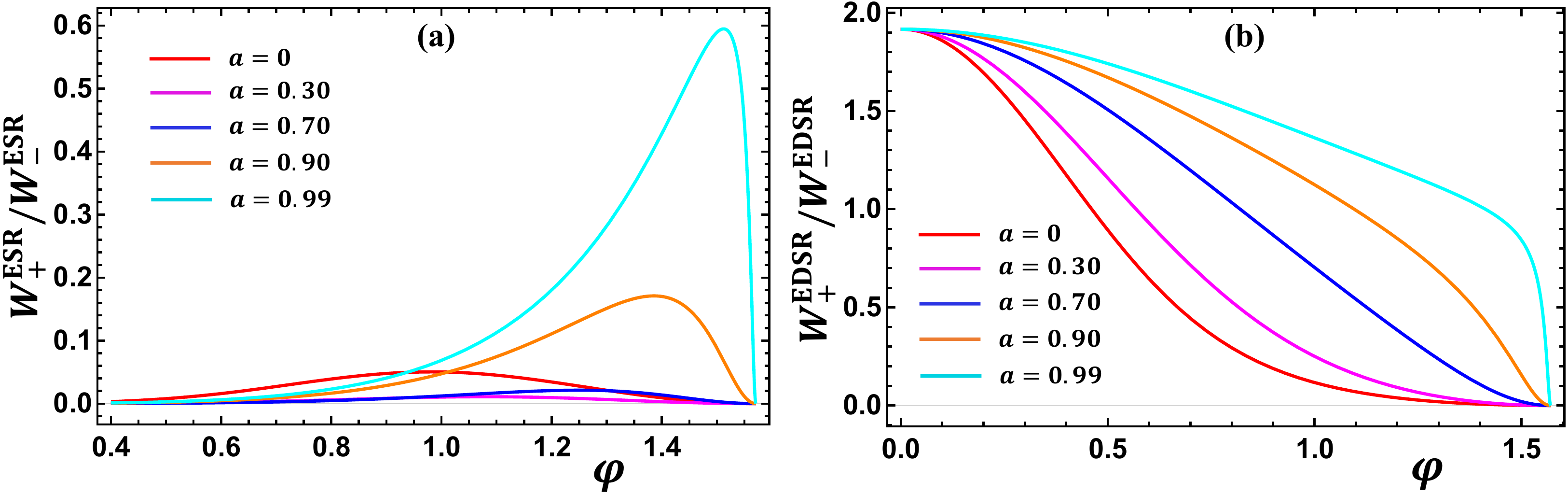}
\caption{\label{residue} Ratio of the oscillator strengths of the two modes for ESR (a) and EDSR (b) as a function of angle $\varphi$, as defined by Eq.~\er{angle}. In both panels, $F_0^a =-0.55= 2F_1^a = 4F_2^a$, $H_0 = aF_0=2H_1 = 4H_2$. The values of $a$ are shown in the legend.}
\end{figure}

\subsection{Zero-field electric-dipole spin resonance in a Fermi liquid} 
\label{sec:edsr}

The electric current is derived in the usual way from the continuity equation for the charge density $\rho=e\text{Tr}\int d^2k \hat n/(2\pi)^2$, see, e.g. Refs.~\onlinecite{nozieres, physkin}. For a spatially non-uniform case, the kinetic equation reads as
\beq
\label{kinetic_edsr}
\frac{\partial \hat{n}}{\partial t} +\frac 12\left\{ \boldsymbol{\nabla}_\bk \hat \ve,  \boldsymbol{\nabla}_\br \hat n\right\} -\frac 12\left\{\boldsymbol{ \nabla}_\br \hat \ve, \boldsymbol{\nabla}_\bk \hat n\right\} + i [\hat{\ve}, \hat{n}] = 0,
\eeq
where $\left\{\hat{\bf a}_1,\hat{\bf a}_2\right\}=\hat{\bf a}_1\cdot\hat{\bf a}_2+\hat{\bf a}_2\cdot\hat{\bf a}_1$,
and $\hat \ve$ and $\hat n$ are still given by Eqs.~\er{sce}-\er{sce_d} and \er{dn_a}-\er{dnsv}, respectively, except for that  their $t$-dependent parts now depend also on $\br$.  Linearizing Eq.~\er{kinetic_edsr}} with respect to deviations from equilibrium, we obtain
\beq
\label{kinetic_edsr_b}
\frac{\partial \delta\hat{n}_{\text{fl}}}{\partial t} +\frac 12\left\{ \boldsymbol{\nabla}_\bk \hat \ve_{\text{eq}},  \boldsymbol{\nabla}_\br  \delta\hat n_{\text{fl}}\right\} -\frac 12\left\{\boldsymbol{ \nabla}_\br \delta\hat \ve_{\mathrm{LF}}, \boldsymbol{\nabla}_\bk \hat n_{\text{eq}}\right\} + i [\hat{\ve}, \hat{n}] = 0.
\eeq
Multiplying the equation above by $-e$, integrating over $\bk$, taking the trace, and taking into account that the trace of any commutator is equal to zero, we obtain the continuity equation $\partial_t\rho+\boldsymbol{\nabla}\cdot {\bf j}=0$ with the current given by
\beq
\label{current}
{\bf j}=-e\text{Tr} \int \frac{d^2p}{(2\pi)^2}\hat\bv_{\text{eq}} \left(\delta \hat n_{\text{fl}}-n_F'\delta\hat\ve_{\mathrm{LF}}\right),
\eeq
where $\hat\bv_{\text{eq}} =\boldsymbol{\nabla}_\bk \hat \ve_{\text{eq}}$ is the equilibrium quasiparticle velocity.
In our case, $\hat \ve_{\text{eq}}$ is given by Eq.~\er{eso} and,  correspondingly, $\hat \bv_{\text{eq}}=\hat\tau_0\hat \xi_0v_F^* \hat \bk+\hat\bv_{\text{SO}}$, where
\bea
\hat\bv_{\text{SO}}=\frac{\lambda_R^*}{2k} \tau_0 (\hat \bk \cdot \hat{\textbf{s}}) (\hat \bk \times \hat{\textbf{z}})\label{vso}
\eea
is the spin-dependent part of the velocity. 

To describe EDSR, we assume that a weak, uniform, oscillatory electric field, $\bE = \bE_0 e^{-i\Omega t}$, is applied in the plane of the graphene layer. Accordingly, the kinetic equation takes the form\footnote{Note that in the presence of an electromagnetic field, $\bk$ entering the kinetic equation is the canonical rather than kinematic momentum\cite{nozieres}}
\beq
\label{kinetic2}
\frac{\partial  \hat n(\bk, t)}{\partial t} - e \bE \cdot \hat\bv_{\text{eq}}
n_F' + i [\hat{\ve} (\bk, t), \hat{n} (\bk, t)] = 0.
\eeq
It is convenient to trace out an equation for the scalar part  of $\delta\hat n_{\text{fl}}$ in Eq.~\er{dn}, i.e., for the function $a(\bk,t)$:
\beq
\frac{\partial  a(\bk, t)}{\partial t} - e v_F^* \bE \cdot \hat \bk n_F' = 0,
\eeq
which gives the spin-independent part of the current. Solving for $a(\bk,t)$ and substituting the result into Eq.~\er{current},
we obtain the Drude part of the conductivity
\beq
\sigma_\text{Drude} = i\frac{(1+F_1^s)}{2} \frac{e^2 v_F^{*2} \nu^*_F}{\Omega}.
\eeq
Parenthetically, we note that because graphene is not a Galilean-invariant system, the renormalized mass, $m^*=k_F/v_F^*$,  is not expressed entirely via the parameter $F_1^s$,\cite{baym,Chubukov:2018} and the Drude weight is renormalized by the interaction.\cite{MacDonald2011} 

Our main interest is the equation for the spin-valley part of the density matrix, which satisfies \beq
\frac{\partial \delta \hat n_{\text{sv}}(\bk, t)}{\partial t} -e \hat\bv_{\text{SO}} \cdot \bE n_F' + i [\hat{\ve} (\bk, t), \hat{n} (\bk, t)] = 0,
\eeq
where $\hat\bv_{\text{SO}}$ is given by Eq.~\er{vso}.
We now use Eqs.~\er{sce}-\er{sce_d} and \er{dn_a}-\er{dnsv} for $\hat \ve(\bk, t)$ and $\hat n(\bk, t)$, respectively, and derive the equations of motion for the components $\bu$ and $\hat M$; see Appendix~\ref{Kin} for technical details. The resonant part of the current is expressed via the solutions of these equations as
\beq
\label{curr_har}
\begin{split}
\textbf{j}_{\text{EDSR}} = \frac{e \lambda_R^* \nu_F^*}{2k_F} \bigg[ \bigg\{ \frac{u_2^{-1} - u_2^1}{2i} f_+ - \frac{u_3^{-1} + u_3^1}{2} f_- \bigg\} \hat{x} - \bigg\{ \frac{u_2^{-1} + u_2^1}{2} f_+ + \frac{u_3^{-1} - u_3^1}{2i} f_- \bigg\} \hat{y} \bigg],
\end{split}
\eeq
where $f_{\pm}$ are defined in Eq.~\er{fh}.
From Eq.~\er{curr_har} we obtain for the resonant part of the conductivity 
\beq
\re\sigma_\text{EDSR}(\Omega) =\sigma_0\lambda_\text{SOC}^* \left[ W^{\rm EDSR}_{+}\delta(\Omega - \Omega_+)
+W^{\rm EDSR}_{-}
 \delta(\Omega - \Omega_-) \right],\; \text{where}\;\sigma_0=\frac{e^2}{4}\frac{\lr^*}{v_F^*k_F}.
\label{resigma}
\eeq
The strengths of oscillators for the two modes are found as $W^{\rm EDSR}_{\pm}=\pm W^{\rm EDSR}(\Omega_\pm)$, where
\beq
W^{\rm EDSR}(\Omega)=
f_+\frac{\Omega^2 - \Omega_a^2}{4\Omega^2_0} \frac{\lambda_R^*}{\lambda_\text{SOC}^*},\;\;
\text{and}\;\Omega_a^2 = h_+ \left[ h \lambda_R^{*2} + \frac{f^2_+-f^2_-}{f_+} \lambda_Z^{*2}\right],
\label{WEDSR}
\eeq 
and $f_{\pm}, h$ and $h_+$ are defined in Eq.~\er{fh}. If both Rashba and VZ types of SOC are present, and for a generic choice of the Landau interaction parameters, the EDSR signal consists of two peaks, as shown in Fig.~\ref{Peaks}e. Because the resonant part of the conductivity in Eq.~\er{resigma} is proportional to $\lr^{*2}$, the EDSR 
 signal is absent without Rashba SOC and weak for small 
 $\lr^*$. This follows already from Eq.~\er{leR}, which shows that electron spins couple to the electric field only due to Rashba SOC.  If $\lr^*\neq 0$ but $\lz^*=0$, then $W^{\rm EDSR}_+$ vanishes [for the same constraints on the Landau parameters described after Eq.~\er{O0}], and the EDSR signal consists of only one peak, see Fig.~\ref{Peaks}f. For $\lambda_Z^*=0$, it is always the spin mode that is active in both ESR and EDSR, while the spin-valley mode remains silent. 
 We see from panels (d) and (e) of Fig.~\ref{Peaks}, as well as from panel (b) in Fig.~\ref{residue}, that, in contrast to the ESR case, the two EDSR peaks are of comparable amplitudes for a wide range of $\lz^*/\lr^*$. 
 
\section{Discussion and conclusions}\label{sec:Conclusions}

In this paper, we predicted that graphene with proximity-induced spin-orbit coupling (SOC) exhibits both a zero-field electron spin resonance (ESR), if probed by an {\em ac} magnetic field, and a zero-field electric-dipole spin resonance (EDSR), if probed by an in-plane {\em ac} electric field. The resonance frequencies are determined by the coupling constants of spin-orbit interaction, as well as by Fermi-liquid (FL) interaction in the spin exchange and spin-valley exchange channels. The most important result of our study is that, if both Rashba and valley-Zeeman (VZ)  types of SOC are present, the ESR and EDSR signals consist of two peaks, which correspond to coupled oscillations of the uniform and the valley-staggered magnetizations.

For graphene  grown on heavy-metal substrates and/or intercalated with heavy-metal atoms, one expects (and does observe) only the Rashba-type SOC. Therefore, our prediction for this type of systems amounts to single ESR and EDSR peaks  at the same frequency, of the order of the Rashba splitting. Depending on the system, $\lambda_R$ varies from 15 to 100 meV, \cite{varykhalov:2008,Marchenko2012,Krivenkov:2017} and thus the resonance frequency straddles the interval from  THz to near infrared range.

Another promising platform is graphene on  transition-metal-dichalcogenide (TMD) substrates. SOC in these systems is also strong, which is evidenced by a strong reduction in the spin-relaxation time, as compared to graphene on light-element substrates. Also,  beatings of Shubnikov-de Haas (ShdH) oscillations observed in high-mobility bilayer graphene on WSe$_2$ provide a direct confirmation of band splitting due to SOC. From these beatings, one estimates the total SOC strength to be $\lambda_{\text{SOC}}=10-15$\,meV, \cite{morpurgo_2016} which places the resonance frequency into the THz range. As to the relative strengths of the Rashba and VZ components of SOC, the situation is more controversial. While experimental studies of weak antilocalization in monolayer graphene on TMD find  VZ SOC to be much stronger than the Rashba one,\cite{wakamura:2018,schonenberger:2018,wakamura:2019} the opposite conclusion is reached in, e.g.,  Refs.~\onlinecite{morpurgo_2015,morpurgo_2016,Yang:2016,Omar:2018}. On the other hand, strong evidence for Rashba SOC being the dominant type in bilayer graphene on WSe$_2$ follows from the dependence of the splitting of the ShdH frequencies on the carrier number density.\cite{morpurgo_2016}
At least partially, therefore, this contradiction may arise from the genuine difference between monolayer samples, used in weak-antilocalization studies of Refs.~\onlinecite{wakamura:2018,schonenberger:2018,wakamura:2019}, and bilayer samples, used in ShdH studies of Ref.~\onlinecite{morpurgo_2016}.  
 Without getting deeper into this discussion, we emphasize that the results of our paper can be used as an independent test for the dominant type of SOC. Indeed, the coupling between the electric field and electron spins is possible only due to Rashba SOC, see Eq.~\er{leR}. Therefore, if the experiment shows no EDSR signal, while the ESR signal contains only a single peak, this would be a clear indication that VZ SOC is the dominant mechanism.  On the contrary, if single peaks (at the same frequency) are observed both by EDSR and ESR, this would indicate that Rashba SOC is the dominant mechanism. Finally, if both ESR and EDSR signals are split into two peaks, this would indicate that the Rashba and VZ types of SOC are of comparable strength. A quantitative analysis of the signal shape should allow one not only to obtain the spin-orbit coupling constants (renormalized by the interaction), but also to extract up to six FL parameters in the $m=0,1,2$ angular momentum channels, which are hard, if at all possible, to be extracted from other types of measurements.

From the experimental point of view, the main issue is how strongly ESR and EDSR peaks are smeared by spin-relaxation  mechanisms, which arise mostly from impurity scattering in the presence of SOC. There are three main types of spin relaxation: i) Mott-like scattering from heavy impurities, which occurs for any kind of the bandstructure; ii) Elliott-Yafet mechanism, which occurs if SOC affects the bandstructure but inversion symmetry is preserved, and iii)  D'yakonov-Perel mechanism, which occurs if SOC affects the bandstructure and inversion symmetry is broken. The contribution of the first two mechanisms to the spin-relaxation rate $\tau^{-1}_{\text{SO}}$ is inversely proportional to the momentum relaxation time ($\tau_p$), whereas the D'yakonov-Perel contribution is linearly proportional to $\tau_p$. Analyzing the dependence of  $\tau^{-1}_{\text{SO}}$ on  $\tau_p$, one can separate the Mott and Elliott-Yafet contributions from the D'yakonov-Perel one.\footnote{In recent literature, the sum of the Mott and Elliott-Yafet contributions is often referred to as just  ``the Elliott-Yafet contribution, which is not correct.} For graphene on TMD, both Rashba and VZ types of SOC contribute to the D'yakonov-Perel mechanism, with VZ contribution being proportional to the intervalley scattering time rather than to $\tau_p.$\cite{cummings:2017}
The results of such an analysis differ from study to study, which indicates that spin relaxation is sample-dependent. For example, Ref.~\onlinecite{morpurgo_2016} does not find any correlation between $\tau_{\text{SO}}$ and $\tau_p$; Ref.~\onlinecite{shi:2017} finds that the D'yakonov-Perel mechanism is the dominant one;  finally, Ref.~\onlinecite{wakamura:2018,wakamura:2019} find that the Mott and/or Elliott-Yafet mechanisms are the dominant one. To be specific, we adopt the last scenario, in which case  Rashba and VZ types of SOC determine the positions of the ESR and EDSR peaks, while their widths are controlled by the Mott and/or Elliott-Yafet mechanisms.
 The most direct estimate for the strength of inversion-symmetry-broken SOC  follows from beating of ShdH oscillations, cited above. On the other hand, Refs.~\onlinecite{wakamura:2018,wakamura:2019}  identify two distinct groups of graphene/TMD systems: with   $\tau^{-1}_{\text{SO}}\sim 0.1-1$\, meV and with $\tau^{-1}_{\text{SO}}>10$\, meV, respectively. The first group contains graphene on monolayer WS$_2$ and on bulk WSe$_2$, while the second group contains graphene on monolayer MoS$_2$ and bulk WS$_2$. Given that $\lambda_{\text{SOC}}\sim 10-15$\, meV,\cite{morpurgo_2015,morpurgo_2016} the quality factor for a resonance on samples from the first group can be as high as 100.

Both ESR and EDSR techniques have their pros and cons. Resistively detected ESR (RDESR) has been observed in graphene with intrinsic (Kane-Mele) SOC.\cite{sichau:2019} For magnetic fields up to 1\,T and with a $g$-factor of $\approx2$ for graphene, we get $g\mu_BB\lesssim 0.1$meV. This is still small compared to typical $\lambda_{\rm R,Z}$ so that one can consider RDESR as a viable technique for observing the zero-field ESR predicted in this paper. On the other hand, our model predicts that the amplitude of the $\Omega_+$-mode (which becomes the spin-valley mode at $\lz^*\rightarrow0$) can be significantly  smaller than the amplitude of the $\Omega_-$ peak (which becomes the spin mode at  $\lz^*\rightarrow0$), see Fig.~\ref{residue}. This may hinder the observation of both mode splitting by ESR. Turning now to EDSR, once Rashba SOC is present, the intensity of the EDSR signal exceeds that of the ESR signal by many orders of magnitude, \cite{rashba:1991,rashbaefros2003,efros2006,shekhter2005,maiti2016,Raines:2021b} which was clearly demonstrated by experiments on 2D quantum wells; see, e.g., Ref.~\onlinecite{Schulte:2005}.  On the other hand, while the EDSR signal is strong, it occurs on top of the Drude tail of the optical conductivity. Given a very light effective mass of charge carriers in graphene, the transport relaxation times are rather short, on the order of a picosecond, even in the highest mobility samples. For a rough estimate, we replace the $\delta$-functions in Eq.~\er{resigma} by Lorentzians of width $\tau^{-1}_{\text{SO}}$, take the best-case scenario, when Rashba and VZ types of SOC are of comparable strength,  i.e., $\lr^*\sim \lz^*\sim \lambda^*_{\text{SOC}}$, assume that $\lambda^*_{\text{SOC}}\tau_p\gg 1$, and also replace all the FL parameters by numbers of order one. Then the amplitude of the EDSR peak can be estimated as $\re\sigma_{\text{EDSR}}\sim e^2 \lambda^2_{\text{SOC}}\tau_{\text{SO}}/\mu$, whereas the Drude tail at the resonance frequency is reduced to $\re\sigma_{\text{Drude}}\sim e^2\mu/\lambda^2_{\text{SOC}}\tau_p$. For the ratio of the two parts of the conductivity we obtain
\beq
\re\sigma_{\text{EDSR}}/\re\sigma_{\text{Drude}}\sim\left(\frac{\lambda_{\text{SOC}}}{\mu}\right)^2\lambda^2_{\text{SOC}}\tau _{\text{SO}}\tau_p.
\label{RD1}
\eeq
For $\lambda_{\text{SOC}}/\mu\sim 0.1$, $\lambda_{\text{SOC}} \sim 10$\,meV, $\tau^{-1}_{\text{SO}}\sim 0.1$\,meV, and $\tau^{-1}_p\sim  1 \text{ps}^{-1}\sim 1$\, meV, we find $\re\sigma_{\text{EDSR}}/\re\sigma_{\text{Drude}}\sim 10$, and thus in this case the EDSR peak should be distinguishable against the Drude background. Next, we consider the worst-case scenario, when Rashba SOC is much weaker than VZ one. In this case, the EDSR frequencies and the frequency $\Omega_a$ in Eq~\er{WEDSR} are of order of $\lz^*$, while $\lr^*$ only enters as an overall factor of the conductivity in Eq~\er{resigma}. In this case, Eq.~ \er{RD1} is replaced by
\beq
\re\sigma_{\text{EDSR}}/\re\sigma_{\text{Drude}}\sim\left(\frac{\lr}{\lz}\right)^2\left(\frac{\lz}{\mu}\right)^2\lz^2\tau _{\text{SO}}\tau_p.
\label{RD2}
\eeq
For  $\lr/\lz=0.1$, $\lz/\mu=0.1$, $\lz=10$\,meV, and the same $\tau^{-1}_{\text{SO}}$ and $\tau^{-1}_p$ as above,  we find $\re\sigma_{\text{EDSR}}/\re\sigma_{\text{Drude}}\sim 10^{-2}$, which makes the observation of the EDSR peak challenging. Therefore, the optimal condition for observing the EDSR peak is $\lr\gtrsim \lz$.

\acknowledgments
We thank H. Bouchiat, C. R. Bowers, L. Glazman, A. Goyal, S. Gueron, L. Levitov, Z. Raines, P. Sharma, and T. Wakamura for stimulating discussions, and   A.  Macdonald for bringing Ref.~\onlinecite{sichau:2019} to our attention. 
This work was supported by the National Science Foundation under Grant No. NSF DMR-1720816 (A.K. and D.L.M.), University of Florida under Opportunity Fund OR-DRPD-ROF2017 (A.K. and D.L.M.), and the Natural Sciences and
Engineering Research Council of Canada (NSERC) Grant
No. RGPIN-2019-05486 (S.M.). D.L.M. acknowledges the hospitality of the Kavli Institute for Theoretical Physics, Santa Barbara, California and of Laboratoire de Physique des Solides, University Paris-Sud, Orsay, France.
The Kavli Institute for Theoretical Physics is supported by  the National Science Foundation under Grant No. NSF PHY-1748958.
\appendix
\section{Electron spin and electric-dipole spin resonances for non-interacting electrons}
\label{app:free}
\subsection{Definitions}\label{Def}
 It will be useful to introduce a complete set of sixteen 4$\times$4 matrices defined as $\hat{\kappa}_{ab} = s_{a} \sigma_{b}$, where $a, b \in 0, x,y,z$, with the convention that
\[
   \hat{\kappa}_{x 0} = 
   \left[ {\begin{array}{cc}
   0 & \sigma_0 \\
   \sigma_0 & 0 \\
   \end{array} } \right],  
   \hat{\kappa}_{y 0} = 
   \left[ {\begin{array}{cc}
   0 & -i \sigma_0 \\
   i \sigma_0 & 0 \\
   \end{array} } \right],
   \hat{\kappa}_{z 0} = 
   \left[ {\begin{array}{cc}
   \sigma_0 & 0 \\
   0 & -\sigma_0 \\
   \end{array} } \right]
\]
and
\[
   \hat{\kappa}_{0 x} = 
   \left[ {\begin{array}{cc}
   \sigma_x & 0 \\
   0 & \sigma_x \\
   \end{array} } \right],  
   \hat{\kappa}_{0 y} = 
   \left[ {\begin{array}{cc}
   \sigma_y & 0 \\
   0 & \sigma_y \\
   \end{array} } \right],
   \hat{\kappa}_{0 z} = 
   \left[ {\begin{array}{cc}
   \sigma_z & 0 \\
   0 & \sigma_z \\
   \end{array} } \right].
\]
The Hamiltonian in the sublattice basis, expressed in terms of $\hat{\kappa}_{a b}$, reads
\beq
\label{ham01}
\hat{H}_0 = v_F(\tau_z \hat{\kappa}_{0x} k_x + \hat{\kappa}_{0y} k_y) + \frac{\lambda_{\rm R}}{2} (\tau_z \hat{\kappa}_{yx} - \hat{\kappa}_{xy}) + \frac{\lambda_{\rm Z}}{2}\tau_z\kappa_{z0}.
\eeq

\subsection{Green's functions}\label{Green}
The single-particle Green's function for the case of Rashba SOC is evaluated as
\beq
\label{gfdef}
\hat{G}^0(i\omega_m, \textbf{k}) = \sum_{\alpha, \beta} \frac{\ket{\alpha \beta} \bra{\alpha \beta}}{i\omega_m + \mu - \ve_{\alpha \beta}(\bk)},
\eeq
where $\omega_m$ is the Matsubara frequency. Explicitly, 
\beq
\label{gf}
\hat{G}^0(i\omega_m, \textbf{k}) = \sum_{\alpha, \beta} \hat{\Omega}_{\alpha, \beta}(\textbf{k}) g_{\alpha, \beta}(i\omega_m, k),
\eeq
where
\beq
\begin{split}
\label{freqindep}
\hat{\Omega}_{\alpha, \beta}(\textbf{k}) =& \frac{1}{4} \bigg\{ \Big[ \hat{\kappa}_{00} + \frac{1 - (\epsilon_{\alpha\beta})^2}{1 + (\epsilon_{\alpha\beta})^2} \tau_z \hat{\kappa}_{zz} \Big] + \frac{2 \epsilon_{\alpha\beta}}{1 + (\epsilon_{\alpha\beta})^2} (\tau_z \hat{\kappa}_{0x} \cos\theta + \hat{\kappa}_{0y} \sin\theta) - 2 \alpha \beta \frac{\epsilon_{\alpha\beta}}{1 + (\epsilon_{\alpha\beta})^2} (\hat{\kappa}_{x0} \sin\theta - \hat{\kappa}_{y0} \cos\theta) \\
+& \alpha \beta \frac{(\epsilon_{\alpha\beta})^2}{1 + (\epsilon_{\alpha\beta})^2} (\tau_z \hat{\kappa}_{yx} - \hat{\kappa}_{xy}) - \alpha \beta \frac{1}{1 + (\epsilon_{\alpha\beta})^2} \sin 2\theta (\tau_z \hat{\kappa}_{xx} - \hat{\kappa}_{zz}) + \alpha \beta \frac{1}{1 + (\epsilon_{\alpha\beta})^2} \cos 2\theta (\tau_z \hat{\kappa}_{yx} + \hat{\kappa}_{xy}) \bigg\},
\end{split}
\eeq
\beq
\begin{split}
\label{chgr}
g_{\alpha, \beta}(i\omega_m, k) =& \frac{1}{i\omega_m + \mu - \ve_{\alpha \beta}(\bk)},
\end{split}
\eeq
where $\epsilon_{\alpha\beta}$ are the dimensionless eigenvalues defined in Eq.~\er{es} and 
all notations are the same as in Sec.~\ref{sec:free_RSOC} of the main text.
The single-particle Green's function for the case of VZ SOC is given by
\bea
\label{gf_vz}
\hat{G}^0(i\omega_m, \bk) &=& \sum_{\alpha, \beta}\hat{\Omega}_{\alpha\beta}(\bk)\frac{1}{i\omega_m + \mu - \ve_{\alpha \beta}(\bk)},\nonumber\\
\hat{\Omega}_{\alpha\beta}(\bk)&=&\frac{1}{4} \Big[ (\hat{\kappa}_{00} + \alpha \beta \tau_z \hat{\kappa}_{z0}) + \alpha (\tau_z \hat{\kappa}_{0x} \cos\theta + \hat{\kappa}_{0y} \sin\theta)+ \beta (\hat{\kappa}_{zx} \cos\theta + \tau_z \hat{\kappa}_{zy} \sin\theta) \Big],\nonumber\\
\eea
where the eigenvalues $\ve_{\alpha\beta}(\bk)$ are given by Eq.~\er{es_vz}.

\subsection{Correlation functions}\label{Corr}
Using the Green's functions presented in Sec.~\ref{Green}, one can calculate the correlation functions of spins and currents, etc. with the help of the Kubo formula. It is convenient to separate out the contributions from the $K$ and $K'$ points.   To this end, we introduce the  correlation functions in the Matsubara domain:
\beq\label{eq:pol}
\Pi^{K/K'}_{X_aX_b}(i\Omega_n) = T\sum_{\omega_n}\int_{\bk\in K/K'}\frac{d^2k}{(2\pi)^2} 
 \text{Tr} \Big[ \hat{X}_a \hat{G}^0 (i\omega_m, \bk) \hat{X}_b \hat{G}^0 (i\Omega_n + i\omega_m, \bk) \Big],
\eeq
where $a,b\in\{x,y,z\}$. The matrix $\hat{X}_a$ denotes the $a^{\text{th}}$ Cartesian component of either the spin $\hat{\mathcal{S}}_a=\hat\tau_0\hat s_a$ or current $\hat{\mathcal{J}}_a=e\hat v_a$  operators, where $\hat\bv$ is the velocity operator given by Eq.~\er{vel}. Any correlation function is the sum of the contributions from the $K$ and $K'$ points.

\subsubsection{Spin susceptibility}
\label{app:free_chi}
The spin susceptibility tensor can be expressed as 
\beq
\chi_{ab}(i\Omega_n)=-\frac{g^2\mu_B^2}{4}\Bigg[\Pi^{K}_{\mathcal{S}_a\mathcal{S}_b}(i\Omega_n)+\Pi^{K'}_{\mathcal{S}_a\mathcal{S}_b}(i\Omega_n)\Bigg].
\eeq 
If only Rashba SOC is present, we find
\bea
\label{spinpolT0_11}
\Pi^{K/K'}_{\mathcal{S}_x\mathcal{S}_x} (i\Omega_n)& =&\Pi^{K/K'}_{\mathcal{S}_y\mathcal{S}_y} (i\Omega_n)= - \frac{\nu_F}{4}\bigg[ \frac{\lambda_{\rm R}^2}{\Omega_n^2 + \lambda_{\rm R}^2} -\frac{\lambda_{\rm R}^2}{8i\Omega_n\mu} \ln \frac{2\mu + \lambda_{\rm R}-i\Omega_n}{2\mu + \lambda_{\rm R}+i\Omega_n}- \frac{\lambda_{\rm R}^2}{8i\Omega_n\mu}\ln \frac{2\mu - \lambda_{\rm R}-i\Omega_n}{2\mu - \lambda_{\rm R}+i\Omega_n} \bigg],\nn\\
\Pi^{K/K'}_{\mathcal{S}_z\mathcal{S}_z} (i\Omega_n) &=& - \frac{\nu_F}{4}
 \bigg[ \frac{2\lambda_{\rm R}^2}{\Omega_n^2 + \lambda_{\rm R}^2} + \frac{i\Omega_n \lambda_{\rm R}^2}{2\mu(\Omega_n^2 + \lambda_{\rm R}^2)} \ln \frac{2\mu - i\Omega_n}{2\mu + i\Omega_n} \bigg],
\eea
where $\nu_F=2\mu/\pi v_F^2$ the total density of states at the Fermi surface. Upon analytic continuation, and adding the $K/K'$ contributions we obtain (for $T=0$ and  $\mu>\lambda_{\rm R}$):
\bea
\text{Im}\chi_{xx} (\Omega) &=& \text{Im}\chi_{yy} (\Omega)=\frac{g^2\mu_B^2\pi\nu_F}{16}\bigg[ \lambda_{\rm R} \delta(\Omega - \lambda_{\rm R}) + \frac{\lambda^2_{\rm R}}{4 \Omega\mu} \Big\{ \Theta \left(\Omega-2\mu - \lambda_{\rm R} \right) + \Theta \left( \Omega-2\mu + \lambda_{\rm R} \right)  \Big\} \bigg],\nn\\
\text{Im}\chi_{zz}(\Omega)&=& \frac{g^2\mu_B^2\pi\nu_F}{16} \bigg[\lambda_{\rm R} \left( 2 - \frac{\lambda_{\rm R}}{2\mu} {\,} \text{ln} {\,} \frac{2\mu + \lambda_{\rm R}}{2\mu - \lambda_{\rm R}} \right)\delta(\Omega - \lambda_{\rm R}) + \frac{\Omega\lambda_{\rm R}^2}{\mu(\Omega^2 - \lambda^2_{\rm R})}\Theta \left( \Omega-2\mu \right) \bigg],\nn\\
\label{chiR}
\eea
As discussed in Sec.~\ref{sec:free_RSOC}, the selection rules indicate that there should be a resonance at $\Omega=\lambda_{\rm R}$, which arises from the transitions between the spin-split branches of the conduction band, and also continua of excitations, starting at  $\Omega = 2\mu\pm\lambda_{\rm R}$ (for the in-plane magnetic field) and $\Omega=2\mu$ (for the out-of-plane magnetic field), due to transitions between the spin-split branches of the conduction and valence bands. All of these features are clearly present in Eq.~\er{chiR}.

A similar calculation for VZ SOC gives:
\bea
\label{vz-pi}
\Pi^{K/K'}_{\mathcal{S}_x\mathcal{S}_x}(i\Omega_n) &=&\Pi^{K/K'}_{\mathcal{S}_y\mathcal{S}_y}(i\Omega_n)= -\frac{\nu_F}{4}
 \frac{2\lambda_{\rm Z}^2}{\Omega_n^2 + \lambda_{\rm Z}^2},\\
\Pi^{K/K'}_{\mathcal{S}_z\mathcal{S}_z}(i\Omega_n) &=&0.
\eea
In real frequencies and on adding the $K$ and $K'$ contributions, we get 
\beq
\text{Im}\chi_{xx}(\Omega)=\text{Im}\chi_{yy}(\Omega)=\frac{g^2\mu_B^2\pi\nu_F}{16}\lambda_{\rm Z}\delta(\Omega-\lz),\;\text{Im}\chi_{zz}(\Omega)=0.
\eeq
In agreement with the selection rules discussed in Sec.~\ref{sec:free_VZ}, only the resonance at $\Omega=\lz$ is present.

The spin susceptibilities for both types of  SOCs  are plotted in Fig.~\ref{NonInt_Spin} of the main text.

\subsubsection{Optical conductivity}
\label{app:free_sigma}
The real part of optical conductivity is related to the current-current  correlation function via
\beq
\label{op}
{\rm Re}\sigma_{xx}(\Omega) = - \frac{v_F^2}{\Omega} [\text{Im} \Pi^{K}_{\mathcal{J}_x\mathcal{J}_x} (\Omega)+\text{Im} \Pi^{K'}_{\mathcal{J}_x\mathcal{J}_x} (\Omega)],
\eeq
where 
$\text{Im} \Pi^{K/K'}_{\mathcal{J}_x\mathcal{J}_x} (\Omega)$ is obtained by analytic continuation of $\Pi^{K/K'}_{\mathcal{J}_x\mathcal{J}_x}(i\Omega_n)$. For the case of Rashba SOC, we get
\bea
\label{pi_44}
\Pi^{K/K'}_{\mathcal{P}_x\mathcal{P}_x}(i\Omega_n) &=& - \frac{e^2\mu}{2\pi v_F^2} \left[ \frac{\Lambda}{2\mu} - 1 - \frac{i\Omega_n}{4\mu} \ln \frac{2\mu - i\Omega_n}{2\mu + i\Omega_n} - \frac{\lambda_{\rm R}^2}{8i\Omega_n\mu} \ln \frac{(2\mu+ \lambda_{\rm R}-i\Omega_n) (2\mu - \lambda_{\rm R}-i\Omega_n)}{(2\mu + \lambda_{\rm R}+i\Omega_n) (2\mu - \lambda_{\rm R}+i\Omega_n)}\right.\nn\\
&&\left.\hspace{2.5cm}+ \frac{i\Omega_n \lambda_{\rm R}^2}{4\mu(\Omega_n^2 + \lambda_{\rm R}^2)} \ln \frac{2\mu + i\Omega_n}{2\mu - i\Omega_n}   \right],
\eea
where $\Lambda$ is the ultraviolet energy-cutoff of the Dirac  spectrum. This results in
\bea
\label{impse}
\text{Re}\sigma_{xx}(\Omega) &=& \frac{e^2}{8} \left[\lambda_{\rm R} \delta(\Omega - \lambda_{\rm R}) {\,} \text{ln} {\,} \frac{2\mu + \lambda_{\rm R}}{2\mu - \lambda_{\rm R}} + \frac{2(\Omega^2-2\lambda_{\rm R}^2)}{\Omega^2-\lambda_{\rm R}^2} \Theta \left( \Omega-2\mu \right)\right.\nn\\
&&\left.\hspace{2.5cm}+ \frac{\lambda_{\rm R}^2}{\Omega^2}\Big\{ \Theta \left( \Omega-2\mu - \lambda_{\rm R} \right) + \Theta \left( \Omega-2\mu + \lambda_{\rm R} \right) \Big\}\right]. 
\eea
By rotational symmetry, $\sigma_{xx}(\Omega)=\sigma_{yy}(\Omega)$. In agreement with the selection rules, the optical conductivity exhibits a resonance at $\Omega=\lr$ and continua starting at $\Omega=2\mu\pm\lr,2\mu$. For $\lr=0$, the last equation is reduced to the universal conductivity of ideal graphene equal to $(e^2/4)\Theta(\Omega-2\mu)$.

For VZ SOC, 
\beq
\label{pi_44z}
\Pi^{K/K'}_{\mathcal{P}_x\mathcal{P}_x}(i\Omega_n) = - \frac{e^2\mu}{2\pi v_F^2}\Bigg[\frac{\Lambda}{2\mu}-1-\frac{i\Omega_n}{8\mu}\ln\frac{2\mu+\lz-i\Omega_n}{2\mu+\lz+i\Omega_n}-\frac{i\Omega_n}{8\mu}\ln\frac{2\mu-\lz-i\Omega_n}{2\mu+\lz-i\Omega_n}\Bigg],
\eeq
and the resulting optical conductivity is 
\bea
{\rm Re}\sigma_{xx}(\Omega)={\rm Re}\sigma_{yy}(\Omega) &=& \frac{e^2}{8} \Big[ \Theta(\Omega - 2\mu - \lambda_{\rm Z}) + \Theta(\Omega - 2\mu + \lambda_{\rm Z}) \Big].
\eea
In contrast to the case of Rashba SOC, there is no resonance at $\Omega=\lz$ but only continua at $\Omega=2\mu\pm\lz$.
For $\lz=0$, the last equation is again reduced to the universal conductivity of ideal graphene. The conductivities for both types of SOC are plotted in Fig.~\ref{NonInt_Cond} of the main text.

\section{Equations of motion}\label{Kin}
In this appendix, we provide technical details of the derivation and solution of the equations of motion for density matrix.  We follow the procedure of Refs.~\onlinecite{shekhter2005,kumar2017,Raines:2021b} to obtain the system of equations for the dynamical variables  $\bu(\bk, t)$, $\bw(\bk, t)$, and $M_{\alpha\beta}(\bk, t)$, parametrizing the spin-valley part of density matrix  \er{dnsv}. We also include an {\em ac} magnetic field, applied along $y$-direction, in our calculation. This corresponds to adding the (time-dependent) Zeeman term, $\delta \hat{\ve}_B = {\Delta_{\mathrm{Z}}^*}\hat{s}_y/2$, to Eq.~\er{sce}.   From here and onwards, we will be suppressing the unity matrices in the spin and valley spaces for brevity.  Substituting $\delta\hat{\ve}(\bk)$ and $\delta\hat{n}(\bk)$ from Eqs.~\er{sce}-\er{sce_d} and \er{dn_a}-\er{dnsv}, respectively, into Eq.~\er{kinetic}, and linearizing with respect to $\bu$, $\bw$, $M_{\alpha\beta}$ and $\Delta_{\mathrm{Z}}$, we obtain an equation for the spin-valley part of the density matrix:
\beq
\label{kinetic_2}
\begin{split}
i \partial_t (\bu \cdot \bxi &+ \bw \cdot \btau + M_{\alpha\beta} \hat{\tau}_\alpha \hat{\xi}_\beta ) = \big[ \delta\hat{\ve}_\text{SO}, 
\bu \cdot \bxi + \bw \cdot \btau 
+ M_{\alpha\beta} \hat{\tau}_\alpha \hat{\xi}_\beta \big] + \big[ \delta \hat{\ve}_B, \delta \hat{\ve}_\text{SO} \big] \\
&+ \Big[ \delta\hat{\ve}_\text{SO}, \frac{1}{4} \text{Tr}' \int\frac{d\theta'}{2\pi}
 \big\{ 
 (\hat{\textbf{s}} \cdot \hat{\textbf{s}}') F^a(
 \vartheta) + (\btau \cdot \btau') 
 G^a(\vartheta) + (\btau \cdot \btau')(\hat{\textbf{s}} \cdot \hat{\textbf{s}}') H(\vartheta) \big\} \\
& \hspace{3cm} \times \big\{ \hta'_0 (\bu' \cdot \bxi') + \bw' \cdot \btau' + M'_{\alpha\beta} \hta'_\alpha \hat{\xi}'_\beta \big\} \Big] ,
\end{split}
\eeq
where $\delta\hat{\ve}_\text{SO} \equiv \delta\hat{\ve}_\text{SO}(\bk)$ is given by Eq.~\er{eso}, $\{\bu, \bw, M_{\alpha\beta}\} \equiv \{\bu(\theta, t), \bw(\theta, t), M_{\alpha\beta}(\theta, t)\}$, $\{\bu', \bw', M'_{\alpha\beta}\} \equiv \{\bu(\theta', t), \bw(\theta', t), M_{\alpha\beta}(\theta', t)\}$, and $\vartheta=\theta-\theta'$. The first, second and third terms on the RHS of Eq.~\er{kinetic_2} can be written as 
\bse
\beq
\begin{split}
\text{First term} =& \Big[ -\frac{\lr^*}{2} 
 \hat{\xi}_3 - \frac{\lz^*}{2} \hta_z \hat{\xi}_1, 
  \bu \cdot \bxi + \bw \cdot \btau
  + M_{\alpha\beta} \hta_\alpha \hat{\xi}_\beta \Big] \\
=& -i \lr^* \big( 
\hat{\xi}_2 u_1 - 
 \hat{\xi}_1 u_2 + M_{\alpha\beta} \hta_\alpha \epsilon_{3\beta \gamma} \hat{\xi}_\gamma \big) \\
& - i \lz^* \big( \hta_z \hat{\xi}_3 u_2 - \hta_z \hat{\xi}_2 u_3 + \hta_y \hat{\xi}_1 w_1 - \hta_1 \hat{\xi}_1 w_2 + M_{\alpha\beta} \hta_z \hta_\alpha \epsilon_{1\beta\gamma} \hat{\xi}_\gamma + M_{\alpha\beta} \epsilon_{3\alpha \gamma} \hta_\gamma \hat{\xi}_\beta \hat{\xi}_1 \big),
\end{split}
\eeq
\beq
\begin{split}
\text{Second term} &= \frac{\Delta_{\mathrm{Z}}}{2} \big[
\hat{\xi}_2 \sin\theta - \hat{\xi}_3 \cos\theta,
 -\frac{\lr^*}{2} 
  \hat{\xi}_3 - \frac{\lz^*}{2} \hta_z \hat{\xi}_1 \big] \\
&= -i\lr^* \frac{\Delta_{\mathrm{Z}}}{2} 
 \hat{\xi}_1 \sin\theta + i\lz^* \frac{\Delta_{\mathrm{Z}}}{2} e^{-i\Omega t} \hta_z (\hat{\xi}_3 \sin\theta + \hat{\xi}_2 \cos\theta),
\end{split}
\eeq
\beq
\begin{split}
\text{Third term}
=& \Big[ -\frac{\lr^*}{2} 
 \hat{\xi}_3 - \frac{\lz^*}{2} \hta_z \hat{\xi}_1, \int\frac{d\theta'}{2\pi}  \left\{
  (\bu' \cdot \hat{\boldsymbol{\xi}}') F^a(\vartheta) + (\bw' \cdot \btau) 
  G^a(\vartheta) + M'_{\alpha\beta} \hta_\alpha \hat{\xi}'_\beta H(\vartheta) \right\} \Big] \\
=& -i \lr^* \int\frac{d\theta'}{2\pi} \Big\{ \big( u_1' 
 \hat{\xi}_2 - u_2' \cos(\vartheta)  
 \hat{\xi}_1 + u_3' \sin(\vartheta) 
  \hat{\xi}_1 \big) F^a(\vartheta) \\
&\hspace{2cm} + \big( M'_{\alpha 1} \hta_\alpha \hat{\xi}_2 - M'_{\alpha 2} \cos(\vartheta) \hta_\alpha \hat{\xi}_1 + M'_{\alpha 3} \sin(\vartheta) \hta_\alpha \hat{\xi}_1 \big) H(\vartheta) \Big\} \\
-i \lz^* \int\frac{d\theta'}{2\pi} & \Big\{ \Big( (u_2' \cos(\vartheta) - u_3' \sin(\vartheta)) \hta_z \hat{\xi}_3 - (u_2' \sin(\vartheta) + u_3' \cos(\vartheta)) \hta_z \hat{\xi}_2 \Big) F^a(\vartheta) \\
&\hspace{1cm} + (w_1' \hta_y \hat{\xi}_1 - w_2' \hta_x \hat{\xi}_1) G^a(\vartheta) + \Big( \frac{M'_{\alpha\beta}}{2i} \hta_z \hta_\alpha [\hat{\xi}_1, \hat{\xi}'_\beta] + M'_{\alpha\beta} \epsilon_{3\alpha\gamma} \hta_\gamma \hat{\xi}'_\beta \hat{\xi}_1 \Big) H(\vartheta) \Big\},
\end{split}
\eeq
\ese
where $\epsilon_{\alpha\beta\gamma}$ is the Levi-Civita symbol.
At the next step, we expand $\{\bu, \bw, M_{\alpha\beta}\}(\theta, t)$ and $\{F^a, G^a, H\}(\theta)$ over a basis of angular harmonics, $\{\bu, \bw, M_{\alpha\beta}\}(\theta, t) = \sum_m e^{i m \theta} \{\bu^{(m)}, \bw^{(m)}, M_{\alpha\beta}^{(m)}\}(t)$ and $\{F^a, G^a, H\}(\theta) = \sum_m e^{i m \theta} \{F_m^a, G_m^a, H_m\}$, to obtain a system of equations for $u_i^{(m)}$, $w_i^{(m)}$ and $M_{\alpha\beta}^{(m)}$. The subset of equations  that couples to an {\em ac} magnetic field reads
\beq
\label{block3}
\begin{split}
\dot{u}_1^{(m)} &= \lr^* \left[f^{(m)}_+ u_2^{(m)} +if^{(m)}_- u_3^{(m)} \right] - \frac{\Delta_{\mathrm{Z}}^* \lr^*}{2} \frac{\delta_{m,1} - \delta_{m, -1}}{2i}, \\
\dot{u}_2^{(m)} &= -\lr^* f^{(m)} u_1^{(m)} + \lz^* \bigg[ h^{(m)}_+ M_{33}^{(m)} -ih^{(m)}_- M_{32}^{(m)} \bigg],\\
\dot{u}_3^{(m)} &= -\lz^* \left[ h^{(m)}_+ M_{32}^{(m)} +ih^{(m)}_- M_{33}^{(m)} \right], \\
\dot{M}_{31}^{(m)} &= \lr^* \left[ h^{(m)}_+ M_{32}^{(m)} +ih^{(m)}_- M_{33}^{(m)} \right], \\
\dot{M}_{32}^{(m)} &= -\lr^* h^{(m)} M_{31}^{(m)} + \lz^* \left[ f^{(m)}_+ u_3^{(m)}-if^{(m)}_- u_2^{(m)} \right] + \frac{\Delta_{\mathrm{Z}}^* \lz^*}{2} \frac{\delta_{m,1} + \delta_{m, -1}}{2}, \\
\dot{M}_{33}^{(m)} &= -\lz^* \left[f^{(m)}_+ u_2^{(m)}+if^{(m)}_- u_3^{(m)} \right] + \frac{\Delta_{\mathrm{Z}}^* \lz^*}{2} \frac{\delta_{m,1} - \delta_{m, -1}}{2i},
\end{split}
\eeq
where, for brevity, we introduced $f^{(m)}\equiv 1+F^a_m$ and $f^{(m)}_\pm\equiv 1+(F^a_{m-1}\pm F^a_{m+1})/2$,
 and similar definitions for $h^{(m)}$ and $h^{(m)}_\pm$. 
 
In the absence of the external magnetic field ($\Delta_{\mathrm{Z}}^*=0$), we obtain a $6\times6$ eigensystem for the frequencies of coupled oscillations in the spin and valley-staggered spin channels. Since the driving term contains only the  $m=\pm1$ harmonics, only the $m=\pm 1$ modes can be excited. Noting that $F^a_{-m}=F^a_{m}$ (and the same for $H$), we only need to solve the system for one of the modes, e.g.,  $m=1$. The $m=-1$ mode is obtained from the $m=1$ by changing $i\rightarrow -i$. For the special case of $\lz^*=0$, when the spin and spin-valley sectors are decoupled, we recover Eqs.~\er{equ} and \er{eqN} of the main text. 
The full result for the frequencies of the $m=\pm 1$ modes are given in Eq.~\er{R+Z} and \er{O0} of the main text. We see that the dynamics of the $m=\pm 1$ modes is controlled by  six Landau parameters: $F^a_{0,1,2}$ and $H_{0,1,2}$.

In EDSR, the system is perturbed by a weak, oscillatory, in-plane electric field. The equations of motion are the same as in Eq.~\er{block3},
except for now the  $\Delta_{\mathrm{Z}}^*$ terms in equations for $\dot{u}^{(m)}_1$, $\dot{M}^{(m)}_{32}$, and $\dot{M}^{(m)}_{33}$ are absent, while the equation for $\dot{u}^{(m)}_2$
acquires 
a driving term\beq
-\frac{e\lr^*}{2k}\bigg[ E_x \frac{\delta_{m,1}-\delta_{m,-1}}{2i} - E_y \frac{\delta_{m,1}+\delta_{m,-1}}{2}\bigg].
\eeq

\bibliography{referenceFile}
\end{document}